\documentclass[aoas,MSNbibl,nameyear,rotating,dvips]{arximspdf}
\usepackage{dcolumn}
\usepackage{multirow}
\usepackage{graphicx}

\doi{10.1214/13-AOAS699} 
\volume{8}
\issue{1}
\pubyear{2014}
\firstpage{48}
\lastpage{73}

\makeatletter
\newcolumntype{d}[1]{D{.}{.}{#1}}
\newproclaim{ass}{Assumption}
\newcommand{\EE}{\mathrm{E}}
\newcommand{\NN}{N}
\newcommand{\NNN}{NN}
\makeatother

\begin{document}
\begin{frontmatter}

\title{Estimation of causal effects using instrumental variables with
nonignorable missing covariates: Application to effect of type of
delivery NICU on~premature infants\thanksref{T1}}
\runtitle{IV with nonignorable missing covariates}

\begin{aug}
\author[A]{\fnms{Fan} \snm{Yang}\corref{}\ead[label=e1]{yangfan@wharton.upenn.edu}\thanksref{m1}},
\author[B]{\fnms{Scott A.} \snm{Lorch}\ead[label=e2]{lorch@email.chop.edu}\thanksref{m2}}
\and
\author[A]{\fnms{Dylan S.} \snm{Small}\ead[label=e3]{dsmall@wharton.upenn.edu}\thanksref{m1}}
\runauthor{F. Yang, S. A. Lorch and D. S. Small}
\affiliation{University of Pennsylvania\thanksmark{m1} and The Children's Hospital of Philadelphia\thanksmark{m2}}
\address[A]{F. Yang\\
D. S. Small\\
Department of Statistics\\
Wharton School\\
University of Pennsylvania\\
400 Jon M. Huntsman Hall\\
3730 Walnut St.\\
Philadelphia, Pennsylvania 19104-6340\\
USA\\
\printead{e1}\\
\phantom{E-mail:\ }\printead*{e3}} 
\address[B]{S. A. Lorch\\
Department of Pediatrics\\
School of Medicine\\
University of Pennsylvania\\
Division of Neonatology\\
The Children's Hospital of Philadelphia\\
Philadelphia, Pennsylvania 19104\\
USA\\
\printead{e2}}
\end{aug}
\thankstext{T1}{Supported by grants from the Agency for Healthcare
Research and Quality and the Measurement, Methodology and Statistics
program of the National Science Foundation.}

\received{\smonth{1} \syear{2013}}
\revised{\smonth{10} \syear{2013}}

%
\begin{abstract}
Understanding how effective high-level NICUs (neonatal intensive care
units that have the capacity for sustained mechanical assisted
ventilation and high volume) are compared to low-level NICUs is
important and valuable for both individual mothers and for public
policy decisions. The goal of this paper is to estimate the effect on
mortality of premature babies being delivered in a high-level NICU vs.
a low-level NICU through an observational study where there are
unmeasured confounders as well as nonignorable missing covariates. We
consider the use of excess travel time as an instrumental variable (IV)
to control for unmeasured confounders. In order for an IV to be valid,
we must condition on confounders of the IV---outcome relationship, for
example, month prenatal care started must be conditioned on for excess
travel time to be a valid IV. However, sometimes month prenatal care
started is missing, and the missingness may be nonignorable because it
is related to the not fully measured mother's/infant's risk of
complications. We develop a method to estimate the causal effect of a
treatment using an IV when there are nonignorable missing covariates as
in our data, where we allow the missingness to depend on the fully
observed outcome as well as the partially o\mbox{bserved} compliance class,
which is a proxy for the unmeasured risk of complications. A~simulation
study shows that under our nonignorable missingness assumption, the
commonly used estimation methods, complete-case analysis and multiple
imputation by chained equations assuming missingness at random, provide
biased estimates, while our method provides \mbox{approximately} unbiased
estimates. We apply our method to the NICU study and find evidence that
high-level NICUs significantly reduce deaths for babies of small
gestational age, whereas for almost mature babies like 37 weeks, the
level of NICUs makes little difference. A sensitivity analysis is
conducted to assess the sensitivity of our conclusions to key
assumptions about the missing covariates. The method we develop in this
paper may be useful for many observational studies facing similar
issues of unmeasured confounders and nonignorable missing data as ours.\vspace*{15pt}
\end{abstract}

%

\end{frontmatter}

\section{Introduction}\label{sec1}
\subsection{Effect of type of delivery NICUs on premature infants}\label{sec1.1}
Premature infants are infants born before a gestational age of 37
complete weeks. Compared to term infants, premature infants have less
time to develop, so that they are at higher risk of death and
complications and often in need of advanced care, ideally in a neonatal
intensive care unit (NICU) [\citet{Proetal}; \citet{Doy};
\citet{Boyetal83}]. There are two types of NICUs---a high-level NICU
is a NICU that has the capacity for sustained mechanical assisted
ventilation and that delivers on average of at least 50 premature
babies per year, whereas a low-level NICU is a unit that does not meet
these requirements. There is literature that shows that delivery at
high-level vs. low-level NICUs is associated with a reduction in
neonatal mortality after controlling for measured confounders [\citet{Phietal07}; \citet{Chuetal10}; \citet{Rogetal04}]. However,
there are unmeasured confounders such as fetal heart tracing test
results and severity of conditions that could bias these results. The
aim of this paper is to use the instrumental variable method along with
a novel method of controlling for nonignorable missing covariates to
obtain unbiased inferences about the effect on neonatal mortality of
premature babies being delivered in a high-level NICU vs. a low-level
NICU. Understanding how effective high-level NICUs are compared to
low-level NICUs is important for both individual mothers deciding
whether to travel a distance to go to a high-level NICU rather than
going to a local low-level NICU, and for public policy decisions about
premature infant care. In the 1970s, a system of perinatal
regionalization was built in most states in which most infants at risk
of complications such as very premature infants would be sent to
regional high-level NICUs [\citet{Lasetal10}]. This
regionalization system has weakened in recent years with more very
premature infants being born in low-level NICUs [\citet{Lasetal10}; \citet{Howetal02}; \citet{RicReeCut95}; \citet{Yeaetal98}]. If high-level NICUs are truly providing considerably better
care for premature babies, then it is valuable to invest resources in
strengthening the perinatal regionalization system, while if high-level
NICUs are providing at best marginal improvements in care, then
strengthening the perinatal regionalization should probably not be a
priority. Additionally, if only certain types of premature babies
benefit from high-level NICUs (e.g., only those below a certain
gestational age), then resources would be best spent on increasing the
rate of high-level NICU delivery for those types of babies. To address
this, we will estimate the effect of high-level NICU delivery for
babies with different characteristics, such as different gestational ages.

The ideal way to assess the effectiveness of high-level NICUs vs.
low-level NICUs would be to randomize pregnant women to deliver at
different level NICUs, but such a study is not ethical or practical. We
instead consider an observational study. We have compiled data on all
babies born prematurely in Pennsylvania between 1995--2005 by linking
birth certificates to death certificates as well as maternal and
newborn hospital records. More than 98\% of the birth certificates
could be linked to the hospital records [\citet{LorBai} for more
details]. We will use the 189,991 records that could be linked in our
analysis. The measured confounders we will consider are gestational
age, the month of pregnancy that prenatal care started (precare) and
mother's education level. If these measured confounders are the only
confounding variables, that is, the only variables that are related to
both level of NICU delivered at and mortality, then we could use
propensity score/matching/regression methods to control for the
confounders. Unfortunately, some key confounders are unmeasured such as
the results of tests like fetal heart tracing which are related to both
how strongly a doctor encourages a~woman to deliver at a high-level
NICU and a baby's risk of mortality. To control for such unmeasured
confounders, we will consider the instrumental variable (IV) method.

\subsection{Instrumental variable approach}\label{sec1.2}
\begin{figure}[b]
\includegraphics{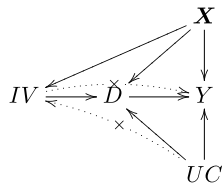}
\caption{This directed acyclic graph shows the assumptions for a valid~IV.
$D$~denotes the treatment, $Y$ the outcome, $\mathbf{X}$ measured
confounders and UC unmeasured confounders. The key assumptions for an
IV are \textup{(i)} the IV affects $D$; \textup{(ii)} the IV does not have a direct effect
on $Y$; \textup{(iii)} the IV is independent of the unmeasured confounders UC
given the measured confounders.}\label{fig1}
\end{figure}

The IV method is widely used in observational studies [\citet{AngKru91}; \citet{Baietal10}]. An instrumental variable (IV)
is a variable that is (i)~associated with the treatment, (ii)~has no
direct effect on the outcome and (iii) is independent of unmeasured
confounders conditional on measured confounders. The relationships
between the~IV, treatment~($D$), outcome ($Y$), measured confounders
($\mathbf{X}$) and unmeasured confounders (UC) are shown in the
directed acyclic graph in Figure~\ref{fig1}. The basic idea of the IV method is
to extract variation in the treatment that is free of the unmeasured
confounders and use this confounder free variation to estimate the
causal effect of the treatment on the outcome. The beauty of the IV
method is that although treatment is not randomly assigned in
observational studies, the method still allows consistent estimation of
the causal effect of a treatment.

The instrumental variable we consider is whether or not the excess
travel time that a mother lives from the nearest high-level NICU
compared to the nearest low-level NICU is less than or equal to 10
minutes; a mother is said to live ``near'' to a high-level NICU if the
excess travel time is $\leq$10 minutes and ``far'' otherwise. Excess
travel time satisfies the first two characteristics of an IV: (i)
association with treatment: previous studies suggest that women tend to
deliver at NICUs near their residential zip code [\citet{LorBai};
\citet{PhiMarLuf93}] and (ii) no direct effect: most women have time
to deliver at both the nearest high-level or other delivery NICU so the
marginal travel time to either facility should not directly affect
outcomes [\citet{LorBai}]. The third assumption needed for excess
travel time to be an IV, that it is independent of unmeasured
confounders conditional on measured confounders, is plausible in that
most women do not expect to have a~premature delivery and hence do not
choose where to live based on distance to a high-level NICU. However,
because high-level NICUs tend to be in certain types of places (e.g.,
in cities) and people living in places with high-level NICUs have
different characteristics from people living far away from high-level
NICUs, for the third IV assumption to hold, we need to condition on
these characteristics that may affect the risk of neonatal death in
these pregnancies. The measured characteristics we are able to
condition on are the month of pregnancy that prenatal care started
(precare), mother's education and gestational age of the baby. We only
have a small number of measured characteristics; for settings where
there are a large numbers of measured characteristics, it is worth
considering Lasso methods to control for the characteristics as in
\citet{ImaRat13}. In previous work [\citet{LorBai}; \citet{GuoChe}], we used
excess travel time as an IV to estimate the effect of high-level vs.
low-level NICUs, but we did not account for the potential nonignorable
missingness of certain measured characteristics. We will develop a
method for accounting for nonignorable missing covariates.

\subsection{Nonignorable missing covariates}\label{sec1.3}

Among the measured confounders, the gestational ages are completely
recorded but some subjects' precare and education level are missing. We
are concerned that the missingness is related with the outcome (death)
and the risks of mother and infant. The information for mother is
usually filled out partly by mother and partly by the nurse or doctor.
If the baby died, the mother may not want to fill out the questionnaire
due to her grief or nurses may not bother the mother to fill out a
questionnaire out of caring for the mother's grief. When the mother or
infant is at high risk of complications, nurses and doctors focus on
this emergency and may ignore recording mother's information.
Consequently, missingness is only plausibly ignorable if we condition
on the outcome (death) and mother's/infant's risk of complications. The
outcome is fully observed but the mother's/infant's risk of
complication is not fully observed. The measured variable gestational
age is a strong predictor of risk but other predictors of risk that are
known to the doctor but not recorded in the data include the results of
fetal heart tracing and the doctor's knowledge about the severity of
mother's and baby's condition. These unmeasured confounders may be
related to the compliance status of the mother. The compliance status
of the mother refers to whether the mother would deliver at a
high-level NICU if she lived near to one (excess travel time $\leq 10$
minutes) and whether she would deliver at a high-level NICU if she
lived far from one (see Section~\ref{sec2.2} for further discussion). If the
mother would always deliver at a high-level NICU regardless of whether
she lives near to one, her compliance status is always taker. If the
mother would only deliver at a high-level NICU if she lives near one,
her compliance status is complier. If a doctor knows that a baby/mother
is at higher risk of complications based on fetal heart tracing or
other knowledge, then the doctor is more likely to recommend the mother
to deliver at a high-level NICU regardless of how near she lives to the
high-level NICU and the mother is more likely be an always taker. Thus,
compliance status is related to unmeasured risk and, consequently, the
missingness of observed variables is likely to be related to compliance
status. Compliance status is only partially observed, for example,
under the assumptions in Section~\ref{sec2.2}, if a mother lives far from a
high-level NICU but still delivers at a high-level NICU, she is an
always taker, but if she lives near a high-level NICU and delivers at a
high-level NICU, she might be an always taker or complier.

Previous literature on IV with missing data has considered missing
outcomes [\citet{FraRub99}; \citet{Meaetal04}; \citet{CheGenZho09}; \citet{SmaChe09};
\citet{LevOMaNor04}]. In this literature, it has
been argued that ignorability of the missing outcome may only be
plausible after conditioning on the covariates \emph{and} the partially
observed compliance status [see~(\ref{equ1})]. Methods have been developed for
estimating causal effects under this ``latent ignorability.'' For missing
covariates rather than missing outcomes, the only work on IV estimation
that we are aware of is \citet{PenLitRag04}, which
assumes missingness of covariates is ignorable conditional on observed
data, but not allowed to depend on compliance behavior. In this paper,
we develop a method for estimation of the causal effect when the
missingness of covariates may depend on the fully observed data as well
as the partially observed compliance behavior.

Generally, if missingness depends only on observed variables, even on
observed outcome, methods like multiple imputation under the assumption
that the data is missing at random (MAR) can provide reasonably good
estimates [\citet{Sch97}]. However, if the missingness of covariates also depends on
partially observed compliance status, multiple imputation methods based
on MAR assumptions may fail to provide valid inference. In this paper,
we will provide a model which allows for missingness to depend on
partially observed compliance status and we use the EM algorithm to
obtain the MLE estimates. We also provide a sensitivity analysis which
allows for missingness to depend on further unobserved confounders
besides compliance status.

Many other observational studies face similar issues of unmeasured
confounding and missing data as ours, and the methods we develop in
this paper may be useful for them. For example, for studying the
comparative effectiveness of two types of drugs, data collected as part
of routine health care practice is often used. Such data may not
contain measurements of important prognostic variables that guide
treatment decisions such as lab values (e.g., cholesterol), clinical
variables (e.g., weight, blood pressure), aspects of lifestyle (e.g.,
smoking status, eating habits) and measures of cognitive and physical
functioning [\citet{Wal96}; \citet{BroSch07}]. To
control for such unmeasured confounders, instrumental variable methods
have been used, for example, the prescribing preference of a patient's
physician for one type of drug vs. the other has been used as an IV
[\citet{KorBau98}; \citet{Broetal06}]. For prescribing
preference to be a valid~IV, it is often necessary to condition on
patient characteristics that differ between different physicians to
account for the possibility that certain physicians tend to see sicker
patients and these physicians may be more likely to prefer one type of
drug than physicians who tend to see less sick patients [\citet{KorBau98}]. However, there is often missing data on some of these
patient characteristics we would like to condition on, in particular,
because the data is collected as part of routine practice rather than
as part of a research study. For example, even if lab tests are always
measured when a lab test is actually administered, since doctors will
only order a lab test for some patients, there will be missing data.
The missingness of lab values might be related to the treatment
decision and outcome and be nonignorable. For example, the decision to
order a lab test is likely related to patient symptoms and/or disease
severity, and we would expect that the probability of a lab test being
ordered depends on what the value of the test would be, if measured,
with unusual values being more likely to be measured [\citet{RoyHen11}]. Thus, comparative effectiveness studies of drugs may need to
consider instrumental variable methods with nonignorable missing
covariates as in our study.

\section{Notation and assumptions}\label{sec2}
\subsection{Notation}\label{sec2.1}
We use the potential outcome approach to define causal effects. Let
$Z_i$ represent the binary IV of infant~$i$; 1 if excess travel time is
less than 10~minutes, which encourages delivery in a high-level NICU; 0
if excess travel time is more than 10 minutes, which does not provide
encouragement of delivery in a high-level NICU. In our data, $56.4\%$
of subjects have excess travel time less than 10 minutes. We use
$\mathbf{Z}$
to denote the vector of IVs for all infants. Let $D_i(\mathbf{z})$ be the
potential binary treatment variable that would be observed for subject
$i$ under IV assignment $\mathbf{z}$. Let $D_i(\mathbf{z})$ be 1
if baby $i$ would be delivered at a high-level NICU under the vector of
$\mathbf{z}$ and 0 if the baby would be delivered at a low-level
NICU. We also let $Y_i(\mathbf{z})$ denote the potential binary outcome,
neonatal death indicator, that would be observed for infant $i$ under IV
assignment $\mathbf{z}$, with $Y_i(\mathbf{z})$ being 1
indicating that the newborn would die in the hospital (neonatal death).
We use $\mathbf{X}_i$ to denote the covariate values for $i$th
subject. The covariates in our study are discrete: infant's gestational
weeks, the month of pregnancy that prenatal care started and mother's
education, namely, 8th grade or less, some high school, high school
graduate, some college, college graduate and more than college. For
simplicity, we include the intercept in $\mathbf{X}_i$. Finally, we
let $R_i^x(\mathbf{z})$ be the binary response indicator of
covariate $x$ under IV $\mathbf{z}$, that is, $R_i^x(\mathbf{z})=1$ if covariate $x$ would be observed for infant $i$ under IV
assignment $\mathbf{z}$, and $R_i^x(\mathbf{z})=0$ if covariate
$x$ would be missing. There is a $R_i^x(\mathbf{z})$ for each
covariate. In the above notation, $D_i(\mathbf{z}),
Y_i(\mathbf{z})$ and $R_i^x(\mathbf{z})$ are all potential outcomes of an
infant. For each infant, depending on the value of $\mathbf{z}$,
one scenario is factual (observed), the other ones are counterfactual
(not observed). We use $D_i$, $Y_i$ and $R^x_i$ to denote observed
treatment received, observed death outcome of infant and the observed
response indicator for covariate~$x$.

\subsection{Assumptions}\label{sec2.2}
We assume the following assumptions hold in our study. The first 5
assumptions are the same as \citet{AngImbRub96}.

\begin{ass}\label{ass1}
Stable unit treatment value assumption (SUVTA),
meaning that a subject's potential outcomes cannot be affected by other
individuals' status.

SUVTA allows us to write $D_{i}(\mathbf{z})$ as $D_{i}(z_i)$,
$Y_{i}(\mathbf{z})$ as $Y_{i}(z_i)$ and $R_i^x(\mathbf{z})=R_i^x(z_i)$. This assumption is plausibly satisfied for our data
since whether a mother delivers at a high-level NICU and her baby's
outcome is unlikely to be affected by other mothers' choice of living
near to a high-level NICU or not.

Based on subjects' compliance behavior, we can partition the population
into four groups:
\begin{equation}\label{equ1}
U_i = \cases{ n, &\quad if $D_i(1)=D_i(0)=0$,
\cr
c, &\quad if $D_i(1)=1$, $D_i(0)=0$,
\cr
a, &\quad
if $D_i(1)=D_i(0)=1$,
\cr
d, &\quad if
$D_i(1)=0$, $D_i(0)=1$,}
\end{equation}
where $n$, $c$, $a$ and $d$ represent never taker, complier, always taker and
defier, respectively. Because $D_i(1)$ and $D_i(0)$ are never observed
jointly, the compliance behavior of a subject is unknown. The parameter
of interest in our study is the complier average causal effect (CACE),
$\mathrm{E}(Y_i(1)-Y_i(0)\mid U_i=c, \mathbf{X}_i=\mathbf{x})$.
\end{ass}

\begin{ass}\label{ass2}
Nonzero average causal effect of $Z$ on $D$. The
average causal effect of $Z$ on $D$, $\mathrm{E}[D_i(1)-D_i(0)]$, is not
equal to zero.

The excess travel time should affect whether mother delivers at a
high-level or low-level NICU due to near NICUs being more convenient,
thus, Assumption~\ref{ass2} is plausible.
\end{ass}

\begin{ass}\label{ass3}
Independence of the instrument from unmeasured
confounders: conditional on $\mathbf{X}$, the random vector [$Y(0)$,
$Y(1)$, $D(0)$, $D(1)$] is \mbox{independent} of~$Z$.

This assumption is plausible for our study because premature delivery
is unexpected for women, so people do not choose where to live based on
the closeness to high-level NICU, especially after controlling for
measured socioeconomic variable such as mother's education level.
\end{ass}

\begin{ass}\label{ass4}
Monotonicity: $D(1)\geq D(0)$.

If a mother is willing to travel to deliver at a high-level NICU when
living 10 or more minutes further to a high-level NICU than a low-level
NICU, she is probably also willing to travel to deliver at a high-level
NICU when living less than 10~minutes further to a high-level NICU than
a low-level NICU.
\end{ass}

\begin{ass}\label{ass5}
Exclusion restrictions among never takers and
always takers:
\[
Y_i(1)=Y_i(0)\qquad\mbox{if }U_i=n\quad\mbox{and}\quad Y_i(1)=Y_i(0)\qquad\mbox{if }U_i=a.
\]

This means that the IV only affects the outcome through treatment and
has no direct effect. In our study, this is plausible because most
women have enough time to make it to either the nearest high-level or
low-level NICU so that marginal travel time should not directly affect
outcomes.
\end{ass}

\begin{ass}\label{ass6}
Nonignorable missingness assumption (missingness
ignorable conditional on compliance class, outcome and fully observed
covariates): suppose the first $k$ covariates of $\mathbf{X}$ are
fully observed and the last $m-k$ covariates have missing values, then
\begin{eqnarray}
P\bigl(R^{X_{i,j}}_i(z)\mid Y_i(z), U_i, \mathbf{X}_i\bigr)=P\bigl(R^{X_{i,j}}_i(z)\mid Y_i(z), U_i, X_{i,1},\ldots,X_{i,k}\bigr)\nonumber
\\
\eqntext{\forall j=k+1,\ldots,m.}
\end{eqnarray}

This is saying that the missingness of covariates precare and mother's
education depends only on neonatal death information, compliance status
of infant and gestational age (fully recorded) as well as the delivery
level of NICU. It is a plausible assumption for our data given the
discussion in Section~\ref{sec1.3}. Identifiability in the simplest setup where
there is only one covariate which is binary under both our nonignorable
missingness assumption and an alternative nonignorable missingness
assumption is discussed in the supplementary material for our paper
[\citet{YanLorSma}].
\end{ass}

\begin{ass}\label{ass7} Exclusion restriction on missing indicator among
never takers and always takers. $R^{X_{i,j}}_i(1)=R^{X_{i,j}}_i(0)$ if
$U_i=n$ and $R^{X_{i,j}}_i(1)=R^{X_{i,j}}_i(0)$ if $U_i=a$.

These are analogous assumptions to \citet{FraRub99}. This
means that the IV has no effect on missingness for never takers and
always takers. We think this assumption is plausible for our data for
the following reasons. We think that the missingness of covariates is
affected by death and the baby's risk of death and complications as
captured by gestational age and compliance class. Since for always
takers and never takers, death is not affected by their level of the IV
$Z$ (this is Assumption~\ref{ass5}) and, additionally, the gestational age and
compliance class are not affected by the level of the IV, the
missingness of covariates for always takers and never takers should not
be affected by the level of the IV.
\end{ass}

\section{Model and estimation}\label{sec3}
We use a general location model [\citet{OlkTat61}; \citet{LitRub02}] for a mixture of continuous and categorical covariate
variables, which could be easily adjusted for cases where covariates
\mbox{variables} are all categorical or all continuous. We consider logistic
models for (i)~treatment assignment given covariates, (ii) outcome in
each compliance class/treatment assignment combination given
covariates, and (iii) missingness in each compliance class/treatment
assignment combination given covariates, and we use a multinomial
logistic model for compliance class.

\textit{Model for covariate}:
suppose that in the $m$ covariates, the first $p$ are categorical and the
remaining $m-p$ are continuous. We assign probability $W_{x_{1},\ldots,x_p}$
to each combination of possible values of those $p$ categorical
covariates variables, where $W_{x_{1},\ldots,x_p}$ are unknown parameters,
and sum up to 1:
\begin{itemize}
\item$(X_{i,1},\ldots,X_{i,p})$ are i.i.d. distributed with
\begin{equation}\label{equ2}
P\bigl((X_{i,1},\ldots,X_{i,p})=(x_{1},\ldots,x_p)
\bigr)=W_{x_{1},\ldots,x_p}\qquad\mbox{where } \sum W_{x_{1},\ldots,x_p}=1.\hspace*{-15pt}
\end{equation}
\item Conditional on $(X_{i,1},\ldots,X_{i,p})=(x_{1},\ldots,x_p)$, we
assume that the continuous covariates random variables
$(X_{i,p+1},\ldots,X_{i,m})$ are multivariate normal with unknown mean
vector $\bolds{\mu}_{x_{1},\ldots,x_p}$, which may depend on the
values of $(x_{1},\ldots,x_p)$, and with unknown common positive definite
covariance matrix $\Sigma$ in order to reduce the number of parameters,
\begin{equation}\label{equ3}
X_{i,p+1},\ldots,X_{i,m}\mid(X_{i,1},\ldots,X_{i,p})=(x_{1},\ldots,x_p)
\mathop{\sim}^{\mathrm{i.i.d.}} \NN_{m-p}(\bolds{\mu}_{x_{1},\ldots,x_p},
\Sigma).\hspace*{-10pt}
\end{equation}
\end{itemize}

\textit{Model for} IV:
\begin{equation}\label{equ4}
P(Z_i=1\mid\mathbf{X}_i=\mathbf{x})=
\frac{\exp (\alpha^{T}\mathbf{x})}{1+\exp{(\alpha^{T}\mathbf{x})}}.
\end{equation}

\textit{Model for compliance class}:
\begin{eqnarray}\label{equ5}
P(U_i=n\mid\mathbf{X}_i=\mathbf{x})&=&
\frac{1}{1+\exp (\delta
_a^{T}\mathbf{x})+\exp{(\delta_c^{T}\mathbf{x})}},
\\
\label{equ6}
P(U_i=c\mid\mathbf{X}_i=\mathbf{x})&=&
\frac{\exp{(\delta
_c^{T}\mathbf{x})}}{1+\exp{(\delta_a^{T}\mathbf{x})}+\exp
(\delta_c^{T}\mathbf{x})},
\\
\label{equ7}
P(U_i=a\mid\mathbf{X}_i=\mathbf{x})&=&
\frac{\exp{(\delta
_a^{T}\mathbf{x})}}{1+\exp{(\delta_a^{T}\mathbf{x})}+\exp
{(\delta_c^{T}\mathbf{x})}}.
\end{eqnarray}

\textit{Model for outcome}:
\begin{equation}\label{equ8}
P\bigl(Y_i(z)=1\mid U_i=u, \mathbf{X}_i=
\mathbf{x}\bigr)=\frac{\exp
{(\beta
_{uz}^{T}\mathbf{x}})}{1+\exp{(\beta_{uz}^{T}\mathbf{x})}}.
\end{equation}
According to Assumption~\ref{ass4}, $\beta_{a0}=\beta_{a1}$ and $\beta
_{n0}=\beta_{n1}$. The quantity of interest is the average treatment
effect for compliers of each covariate level, which is estimated by
$\mathrm{E}(Y(1)-Y(0)\mid U=c, \mathbf{X}=\mathbf{x}) =
\frac
{1}{1+\exp{(\beta_{c0}^{T}\mathbf{x})}}-\frac{1}{1+\exp{(\beta
_{c1}^{T}\mathbf{x})}}$.

\textit{Model for missingness indicators}:
\begin{eqnarray}\label{equ9}
&& P\bigl(R^{X_{i,j}}_i(z)=1\mid Y_i(z)=y,
U_i=u, \mathbf{X}_{i,1,\ldots,k}=\mathbf{x}_{1,\ldots,k}
\bigr)
\nonumber\\[-8pt]\\[-8pt]
&&\qquad =\frac{\exp{(\theta
_{j,u}^T\mathbf{x}_{1,\ldots,k}+\gamma_{j,u}I_{y=1}}+\eta
_{j,u}I_{z=1})}{1+\exp{(\theta_{j,u}^T\mathbf{x}_{1,\ldots,k}+\gamma
_{j,u}I_{y=1}}+\eta_{j,u}I_{z=1})},\nonumber
\end{eqnarray}
where $ j=k+1,\ldots,m$. Based on Assumption~\ref{ass7}, $\eta_{j,a}=\eta_{j,n}=0$,
$\forall j=k+1,\ldots,m$.

Under the models (\ref{equ2})--(\ref{equ9}), we seek to maximize the likelihood of the
joint distribution of $X$, $Z$, $U$, $Y$, $R$. If we know the compliance classes
and the missing covariates for each subject, we can get the MLE of
parameters involved in those models easily. Based on this idea, we are
going to use the EM algorithm.

\subsection{EM algorithm}\label{sec3.1}
For simplicity, we are going to present the EM algorithm for the case
where all the covariates are categorical and that there are 4
covariates (including intercept) with only the first two completely
observed, which is the case of our data. The EM algorithm can be easily
extended to other scenarios. The first covariate is the intercept, and
we further assume that the other three covariates are ordered
categorical with $q_2$, $q_3$, $q_4$ levels, respectively. For a nominal
categorical variable, we can use indicator functions for each category,
which the following algorithm could be easily adjusted for.

Let $\NN_{r_3,r_4, x_2,x_3,x_4,u,z,y}$ be the number of cases where
$R^{X_{3}}_i=r_3$, $R^{X_{4}}_i=r_4$, $X_{i,2}=x_2$, $X_{i,3}=x_3$,
$X_{i,4}=x_4$, $Z_i=z$, $Y_i=y$, $U_i=u$. Notice that $X_{i,1}=1$,~$\forall i$.
Those numbers are only partially observed, however, if they are known,
the complete data log likelihood is
\begin{eqnarray*}
l_c&=&\sum_{r_3,r_4, x_2,x_3,x_4,u,z,y}\NN_{r_3,r_4,
x_2,x_3,x_4,u,z,y}
\\
&&\hspace*{68pt}{}\times \bigl(\log(W_{x_2,x_3,x_4})+\log\bigl(P\bigl(Z_i=z\mid X_i=(1,x_2,x_3,x_4)\bigr)\bigr)
\\
&&\hspace*{82pt}{}+\log\bigl(P\bigl(U_i=u\mid X_i=(1,x_2,x_3,x_4)
\bigr)\bigr)
\\
&&\hspace*{82pt}{} +\log\bigl(P\bigl(Y_i=y\mid Z_i=z,U_i=u,X_i=(1,x_2,x_3,x_4)
\bigr)\bigr)
\\
&&\hspace*{82pt}{}+\log\bigl(P\bigl(R_i^{X_{i,3}}=r_3\mid
Z_i=z, Y_i=y, U_i=u, X_{i,2}=x_2
\bigr)\bigr)
\\
&&\hspace*{82pt}{}+\log\bigl(P\bigl(R_i^{X_{i,4}}=r_4\mid
Z_i=z, Y_i=y, U_i=u, X_{i,2}=x_2
\bigr)\bigr)\bigr).
\end{eqnarray*}

Once we know $\NN$, the MLE estimates of the logistic models in (\ref{equ4})--(\ref{equ9}) are
standard, and the MLE for $W_{x_2,x_3,x_4}\propto
\NN_{\ldots,x_2,x_3,x_4,\ldots}$, where $\NN_{\ldots,x_2,x_3,x_4,\ldots}$ is defined to be
$\sum_{r_3,r_4, u,z,y}\NN_{r_3,r_4, x_2,x_3,x_4,u,z,y}$.

In the E-step, conditional on observed data and parameters' estimates
obtained through the previous step, we can get the expected values for
$\NN_{r_3,r_4, x_2,x_3,x_4,u,z,y}$.

From the observed data, we can get the following counts:
\begin{enumerate}[4.]
\item $\NNN_{x_2,x_3,x_4,d,z,y}$, which denotes the number of cases that
$X_{i,3}$, $X_{i,4}$ are both observed and that $X_{i,2}=x_2$,
$X_{i,3}=x_3$, $X_{i,4}=x_4$, $D_i=d$, $Z_i=z$, $Y_i=y$.
\item$\NN3_{x_2,x_4,d,z,y}$, which denotes the number of cases that only
$X_{i,3}$ are unobserved and that $X_{i,2}=x_2$, $X_{i,4}=x_4$, $D_i=d$,
$Z_i=z$, $Y_i=y$.
\item$\NN4_{x_2,x_3,d,z,y}$, which denotes the number of cases that only
$X_{i,4}$ are unobserved and that $X_{i,2}=x_2$, $X_{i,3}=x_3$, $D_i=d$,
$Z_i=z$, $Y_i=y$.
\item$NB_{x_2,d,z,y}$, which denotes the number of cases that
$X_{i,3}, X_{i,4}$ are both missing and that $X_{i,2}=x_2$, $D_i=d$,
$Z_i=z$, $Y_i=y$.
\end{enumerate}
Further, let $P_{r_3,r_4, x_2,x_3,x_4,u,z,y}$ be the probability of a
subject having a case where $R^{X_{3}}_i=r_3$, $R^{X_{4}}_i=r_4$,
$X_{i,2}=x_2$, $X_{i,3}=x_3$, $X_{i,4}=x_4$, $Z_i=z$, $Y_i=y$, $U_i=u$ which are
calculated based on models (\ref{equ2})--(\ref{equ9}). Then we can get the expected values
for each $\NN_{r_3,r_4, x_2,x_3,x_4,u,z,y}$, for example,\[
\mathrm{E}\NN_{1,1,x_2,x_3,x_4,a,1,y}=\NNN_{x_2,x_3,x_4,1,1,y}\frac
{P_{1,1,x_2,x_3,x_4,a,1,y}}{P_{1,1,x_2,x_3,x_4,a,1,y}+P_{1,1,x_2,x_3,x_4,c,1,y}}
\nonumber.
\]

To save space, all the formulas to update each $\NN_{r_3,r_4,
x_2,x_3,x_4,u,z,y}$ are given in \hyperref[app]{Appendix}. By iteratively finding the
E-step estimate of $\NN$ and maximizing the expected value of the complete
data log likelihood in the M-step until the algorithm converges, we
obtain estimates of the parameters in models (\ref{equ2})--(\ref{equ9}). The $R$ code for
the algorithm to analyze our data is in the supplementary materials for
our paper [\citet{YanLorSma}].

\section{Simulation}\label{sec4}
In this section we conduct simulation studies to estimate the complier
average causal effect in the simplest context where there is only one
covariate, the values of which could only be 0, 1. We consider the
following three scenarios under Assumptions~\ref{ass1}--\ref{ass7}: (1)~covariate is
missing completely at random; (2)~covariate is missing at random,
meaning that the missingness does not depend upon the unobserved data,
for example, does not depend on latent compliance status; (3)~missing
mechanism for covariate is nonignorable: the missingness of covariate
can depend on not only the observed outcome $Y$, treatment assignment~$Z$,
but also latent compliance status~$U$.

In each scenario, we are going to apply the following three estimation
methods and compare their results: (1)~Complete-case analysis, which
provides unbiased estimates when the missing mechanism of the data is
missing completely at random. (2) The estimates using multiple
imputation by chained equations [conducted by MICE, see \citet{VanGro}] which gives valid estimates when data are
missing at random. (3) Our method, which is designed to deal with
nonignorable missingness of covariates.

In the single covariate case, the models described in Section~\ref{sec3} can be
represented simply by the following set of parameters: $W_u$, which is
$P(U_i=u)$; $M_{u}$, which is $P(X_i=1\mid U_i=u)$; $\xi_{x}$, which
represents $P(Z_i=1\mid X_i=x)$; $\theta_{zux}$, which denotes
$P(Y_i(z)=1\mid U_i=u, X_i=x)$; and $\rho_{yzu}$, which are parameters
for missingness indicators $P(R_i(z)=1\mid Y_i=y, U_i=u )$, where
$R_i=0$ if the covariate for the $i$th subject is missing. $\theta
_{1c1}-\theta_{0c1}$ and $\theta_{1c0}-\theta_{0c0}$ are the
corresponding compliers' average causal effect for subjects with $X$
being 1 and 0, respectively.

In all three scenarios, the parameters other than the ones in the
missingness model are arbitrarily chosen and fixed as follows:
\begin{eqnarray*}
W_n&=&0.2,\qquad W_a=0.375,\qquad M_{n}=0.5,\qquad M_{a}=0.25,\qquad M_{c}=0.8,
\\
\xi_{1}&=&0.4,\qquad \xi_{0}=0.6,
\\
\theta_{1n1}&=&0.5,\qquad \theta_{1n0}=0.3,\qquad \theta_{0a1}=0.8,\qquad \theta _{0a0}=0.7,
\\
\theta_{1c1}&=&0.7,\qquad\theta_{1c0}=0.45,\qquad\theta_{0c1}=0.45,\qquad \theta _{0c0}=0.3.
\end{eqnarray*}
The missingness parameters in each scenario are described below; the
values for $\rho$'s are chosen to generate $12\%$ missingness for
covariate (the same missing rate as in the NICU study), and satisfy the
exclusion restriction for missing indicator, which implies that $\rho
_{y0a}=\rho_{y1a}$ and $\rho_{y0n}=\rho_{y1n}$. In the first case, the
missingness parameters $\rho$'s are the same for all possible outcomes,
IV levels as well as compliance classes, thus, the covariate is missing
completely at random; in the second case, the missing rates are
different for different outcomes and IV levels, however will not be
affected by partially observed compliance status, so that the
missingness will not depend on unobserved data, which is a case of missing
at random; in the last case, besides outcome and IV, the compliance
status also plays a role in deciding the probability of missingness,
and the values of $\rho$'s are chosen so that even the largest effect
of compliance status on missingness is still moderate ($\rho
_{11a}-\rho
_{11n}=0.25$) and realistic:
\begin{enumerate}
\item Missing completely at random
\[
\rho_{11n}=\rho_{01n}=\rho_{10a}=\rho_{00a}=
\rho _{11c}=\rho_{01c}=\rho_{00c}=\rho_{10c}=0.88.
\]
\item Missing at random
\begin{eqnarray*}
\rho_{11n}&=&\rho_{10n}=\rho_{10c}=0.88,\qquad \rho_{10a}=\rho_{11a}=\rho _{11c}=0.78,
\\
\rho_{01n}&=&\rho_{00n}=\rho_{00c}=0.94,\qquad \rho_{00a}=\rho_{01a}=\rho _{01c}=0.97.
\end{eqnarray*}
\item Nonignorable missingness
\begin{eqnarray*}
\rho_{11n}&=&\rho_{10n}=0.75,\qquad \rho_{01n}=\rho_{00n}=0.8,\qquad \rho _{10a}=\rho _{11a}=1,
\\
\rho_{00a}&=&\rho_{01a}=0.95,\qquad\rho_{11c}=0.8,\qquad \rho_{01c}=0.9,
\\
\rho _{00c}&=&0.83,\qquad \rho_{10c}=0.97.
\end{eqnarray*}
\end{enumerate}
We simulated 500 data sets for each scenario described above with each
simulated data set containing 5000 subjects. Under the above setup, the
CACE for subjects with covariate being 1 is 0.25, whereas the CACE for
subjects with covariate\vadjust{\goodbreak} 0 is~0.15. Table~\ref{tab1} shows the means and standard
deviations for the estimates of CACE across 500 simulated data sets
using the EM algorithm based on our nonignorable missingness
assumption, the complete-case estimates and multiple imputation
estimates using MICE for each missingness mechanism. The corresponding
bias in percentage is given in parentheses.

%
\begin{table}[t]
\tabcolsep=0pt
\caption{Simulation results under MCAR, MAR and nonignorable missing mechanism}\label{tab1}
{\fontsize{8.5}{11}\selectfont{\begin{tabular*}{\tablewidth}{@{\extracolsep{\fill}}@{}lc c c c c c@{}}
\hline
&\multicolumn{2}{c}{\textbf{EM(NI)}} &\multicolumn{2}{c}{\textbf{Complete case}} &\multicolumn{2}{c}{\textbf{MICE}}\\[-6pt]
&\multicolumn{2}{c}{\hrulefill} &\multicolumn{2}{c}{\hrulefill} &\multicolumn{2}{c}{\hrulefill}\\
\textbf{CACE}&\textbf{Mean} & \textbf{SD} & \textbf{Mean} & \textbf{SD} & \textbf{Mean} & \textbf{SD}\\
\hline
\multicolumn{7}{@{}c@{}}{MCAR}\\
$\theta_{1n1}-\theta_{0n1}=0.250$&0.250 (0.00\%)&0.027&0.249 (0.40\%)\phantom{0}&0.028&0.248 (0.80\%)&0.028\\
$\theta_{1n0}-\theta_{0n0}=0.150$&0.149 (0.67\%)&0.095&0.148 (1.33\%)\phantom{0}&0.096&0.154 (2.67\%)&0.095
\\[6pt]
\multicolumn{7}{@{}c@{}}{MAR}\\
$\theta_{1n1}-\theta_{0n1}=0.250$&0.250 (0.00\%)&0.027&0.221 (11.60\%)&0.029&0.246 (1.60\%)&0.028\\
$\theta_{1n0}-\theta_{0n0}=0.150$&0.147 (2.00\%)&0.097&0.113 (24.67\%)&0.096&0.160 (6.67\%)&0.097
\\[6pt]
\multicolumn{7}{@{}c@{}}{Nonignorable}\\
$\theta_{1n1}-\theta_{0n1}=0.250$&0.250 (0.00\%)&0.027&0.188 (24.80\%)&0.029&0.234 (6.40\%)&0.029\\
$\theta_{1n0}-\theta_{0n0}=0.150$&0.148 (1.33\%)&0.093&0.089 (40.60\%)&0.096&\phantom{0}0.221 (47.33\%)&0.084\\
\hline
\end{tabular*}}}
\end{table}

From Table~\ref{tab1} we see that when data is missing completely at random, all
three methods provide unbiased estimates. In the second scenario when
the missingness depends on observed data, we can no longer obtain
unbiased estimates from complete-case analysis, whereas both our EM
algorithm for nonignorable missingness and MICE designed for data
missing at random still provide reasonable estimates as we expected.
However, when the missingness of covariates depends not only on the
observed outcome, but also on the partially observed compliance status,
simply using the complete cases or assuming missing at random to impute
missing covariates based on the observed data gives us biased estimates
of CACE. The complete-case analysis provides biased estimates due to
the fact that it is actually estimating $\mathrm{E}(Y_i(1)-Y_i(0)\mid
U_i=c, R_i=1)$, which is generally different from $\mathrm
{E}(Y_i(1)-Y_i(0)\mid U_i=c)$ when the data is not missing completely
at random. Imputation based on missingness at random is actually
imputing $X$ as if the missing mechanisms for compliers and always takers
assigned to treatment are the same, and that for compliers and never
takers assigned to control are the same. When this is not the case, the
imputation estimates are biased.

From our simulation study, we can see that even if the missingness rate
of a covariate is low ($12\%$), and the compliance class has only a
moderate effect on the missingness, it is still important and necessary
to model the effect of compliance class on missingness in the analysis,
otherwise the results could be significantly biased.

\section{Application to NICU study}\label{sec5}
The data describes 189,991 babies born prematurely in Pennsylvania
between 1995--2005. These premature babies are the ones whose
gestational ages are between 23 and 37 weeks. The outcome variable we
are interested in is neonatal death of babies, which refers to death
during the initial birth hospitalization; we use $Y_i$ to represent the
outcome of the $i$th baby in the data set, with $Y_i$ being 1
indicating the death of baby~$i$. We view infants that are delivered in a
high-level NICU as the treatment group, whereas the ones that are
delivered in a low-level NICU are the control group. Let $D_i$ equal 1
if the $i$th baby is delivered in a high-level NICU, 0 if in a
low-level NICU. The instrumental variable we consider is whether or not
the mother's excess travel time that a mother lives from the nearest
high-level NICU compared to the nearest low-level NICU is less than or
equal to 10 minutes. As we discussed in Section~\ref{sec2.2}, mother's excess
time is a plausible IV in our study which satisfies the IV Assumptions~\ref{ass1}--\ref{ass7} in Section~\ref{sec2.2}. We use $Z_i$ to denote the IV value for the
$i$th baby, with $Z_i$ being 1~indicating that the excess travel
time is less than 10 minutes. The measured confounders $\mathbf{X}_i$ for baby $i$ are baby's gestational age, the month of pregnancy
that prenatal care started and mother's education. We also include an
intercept in $\mathbf{X}_i$.

%
\begin{sidewaystable}
\tabcolsep=12pt
\textwidth=\textheight
\tablewidth=\textwidth
\caption{Percentages of always takers, compliers and never takers in \%}\label{tab2}
\begin{tabular*}{\tablewidth}{@{\extracolsep{\fill}}@{}lcc c c d{2.1} c d{2.1} c@{}}
\hline
\multirow{2}{44pt}{\textbf{Gestational age}} & \textbf{Precare} & \multirow{2}{38pt}{\textbf{Mother's education}}
&\multicolumn{2}{c}{\textbf{Percentage of always takers}} &\multicolumn{2}{c}{\textbf{Percentage of compliers}}
&\multicolumn{2}{c@{}}{\textbf{Percentage of never takers}}\\[-3pt]
& & & \multicolumn{2}{c}{\hrulefill} &\multicolumn{2}{c}{\hrulefill} &\multicolumn{2}{c@{}}{\hrulefill}
\\
&&&\textbf{Estimate}& \textbf{95\% CI} & \multicolumn{1}{c}{\textbf{Estimate}} & \textbf{95\% CI} & \multicolumn{1}{c}{\textbf{Estimate}} & \textbf{95\%CI}\\
\hline
24&2&High School&87.2& [86.3, 88.0]&4.4&[3.6, 5.1]&8.4&[7.8, 9.0]\\
24&4&High School&87.4& [86.5, 88.3]&4.8&[3.9, 5.6]&7.8&[7.3, 8.5]\\
24&2&College&92.0& [91.5, 92.6]&2.7&[2.2, 3.2]&5.3&[4.8, 5.7]\\
24&4&College&92.2& [91.5, 92.7]&2.9&[2.4, 3.5]&4.9&[4.5, 5.3]\\
30&2&High School&59.4& [57.9, 60.2]&20.0&[18.4, 21.4]&21.0&[20.2, 21.8]\\
30&4&High School&58.9& [57.7, 60.3]&21.6&[19.9, 23.1]&19.5&[18.8, 20.3]\\
30&2&College&71.1& [70.1, 72.0]&14.0&[12.8, 15.1]&15.0&[14.3, 15.6]\\
30&4&College&70.9& [69.8, 72.1]&15.1&[13.8, 16.4]&13.9&[13.3, 14.6]\\
37&2&High School&17.4& [17.1, 17.7]&54.2&[53.6, 54.9]&28.4&[27.9, 28.8]\\
37&4&High School&17.0& [16.5, 17.4]&57.3&[56.6, 58.0]&25.8&[25.2, 26.3]\\
37&2&College&26.5& [26.0, 26.9]&48.0&[47.2, 48.8]&25.6&[25.1, 26.0]\\
37&4&College&25.9& [25.2, 26.6]&50.8&[49.9, 51.8]&23.3&[22.7, 23.9]\\
\hline
\end{tabular*}
\end{sidewaystable}

In this data set, all variables mentioned above are fully observed
except the month of pregnancy that prenatal care started and mother's
education level. The missing rates for those two covariates are 10.3\%
 and 2.3\%, respectively. We did Chi-Square tests of independence to
test whether the missingness of those two covariates depends on outcome~$Y$. The $p$-values are both below $10^{-15}$, strong evidence that
missingness depends on the outcome. We also did logistic regression to
test whether the missing indicators also depend on the observed risk
characteristic of gestational age given the outcome of neonatal death.
The results show that gestational age has a significant negative
association with the missingness of those two covariates even
conditional on outcome ($p$-values are both below $10^{-15}$). Since we
have strong evidence that the missingness depends on observed risk
characteristics, we believe that the missingness should also depend on
unobserved risk characteristics which are reflected in compliance status.

Table~\ref{tab2} describes the estimated proportions of each compliance class---always takers, compliers and never takers---for some typical
combinations of covariates. There is a clear trend that as the
gestational ages get larger, the proportion of always takers gets
smaller, and the proportions of the other two compliance classes get
larger. A reasonable explanation for this phenomenon is that the
gestational age is a strong predictor for the risk of complications as
well as death---the smaller the baby is, the higher risk the baby and
mother have. For babies or mothers at higher risk of complications or
death, doctors are more likely to encourage them to go to a high-level
NICU no matter if the mother lives near one or not, that is, those
mothers are more likely to be always takers. Notice that from the fit
of our model, there is a substantial proportion of never takers,
although it may be surprising that people would choose to bypass a
high-level NICU for a low-level NICU (i.e., be a never taker). Choice
of hospital is driven by a number of factors, including where a
patient's physician practices, the general view of the hospital by a
specific community of patients, and what family or friends believe
about a hospital. There are families who choose to deliver at smaller
hospitals regardless of where they live and their illness severity.
This may be because some families are suspicious of academic hospitals,
which make up the majority of high-level NICUs, and would rather travel
to deliver at a community hospital even if the hospital has fewer
resources to care for them.

%
\begin{table}
\tabcolsep=0pt
\caption{Estimates of outcome model for compliers}\label{tab3}
\begin{tabular*}{\tablewidth}{@{\extracolsep{\fill}}@{}lcc c c@{}}
\hline
\textbf{Parameters} & \textbf{Intercept} & \textbf{Gestational age} & \textbf{Precare} & \textbf{Mother's education}\\
\hline
$\beta_{c1}$&1.400 (1.617)&$-$0.153 (0.043)&0.091 (0.063)&$-$0.522 (0.118)\\
$\beta_{c0}$&9.450 (1.274)&$-$0.395 (0.042)&0.144 (0.055)&$-$0.315 (0.113)\\
\hline
\end{tabular*}
\end{table}

%
\begin{table}[b]
\tabcolsep=0pt
\caption{CACE with different covariate values}\label{tab4}
\begin{tabular*}{\tablewidth}{@{\extracolsep{\fill}}@{}lcc d{2.3} d{3.11}@{}}
\hline
\textbf{Gestational age} & \textbf{Precare} & \textbf{Mother's education} & \multicolumn{1}{c}{\textbf{CACE}} & \multicolumn{1}{c}{\textbf{95\% confidence interval}}\\
\hline
24&2&High School&-0.296& [-0.429, -0.137]\\
24&4&High School&-0.343& [-0.490, -0.162]\\
24&2&College&-0.192& [-0.349, -0.064]\\
24&4&College&-0.230& [-0.421, -0.077]\\
30&2&High School&-0.032& [-0.043, -0.017]\\
30&4&High School&-0.040& [-0.056, -0.023]\\
30&2&College&-0.019& [-0.033, -0.008]\\
30&4&College&-0.024& [-0.043, -0.009]\\
37&2&High School&0.001& [-0.001, 0.002]\\
37&4&High School&0.001& [-0.001, 0.002]\\
37&2&College&0.000& [-0.001, 0.001]\\
37&4&College&0.000& [-0.002, 0.001]\\
\hline
\end{tabular*}
\end{table}

Table~\ref{tab3} shows the estimates of parameters in outcome model for
compliers, which are the parameters to estimate the CACE, $\mathrm
{E}(Y(1)-Y(0)\mid U=c, X=x) = \frac{1}{1+\exp{(\beta
_{c0}^{T}x)}}-\frac
{1}{1+\exp{(\beta_{c1}^{T}x)}}$. The standard errors for the
corresponding parameters are provided in parentheses; the standard
errors are estimated through bootstrap using 1000 re-samples. From the
estimates for the outcome model, we see that larger gestational age and
higher mother's education level are related to low death rate, and that
for the mothers who started prenatal care late, the baby is at more
risk of death.

Table~\ref{tab4} shows the estimated CACE of delivering at high-level NICU vs.
low-level NICU for various combinations of the measured covariates.
High-level NICUs substantially reduce the probability of death for very
premature babies. For example, for an infant of gestational age 24
weeks, whose mother started prenatal care in the second month of
pregnancy and has a high school education, being delivered in a
high-level NICU will reduce the probability of death by 0.296, with a
95\% confidence interval of $-$0.429 to $-$0.137. The effect of high-level
NICUs is less for less premature babies; when the baby's gestational
age is about 37 weeks, the high-level NICU has almost no effect on
mortality. This is plausible since a 37-week baby is almost mature and
is at less risk and, consequently, the type of delivery NICU may not
matter much.

Using our method, the estimated CACE weighted by the probability of
each combination of the measured covariates is $-$0.010, with a $95\%$
confidence interval [$-$0.014, $-$0.006]; and the estimated CACE weighted
by the number of compliers in each combination of the measured
covariates is $-$0.002, with a $95\%$ confidence interval [$-$0.004,
$-$0.001]. Thus, our analysis shows that high-level NICU significantly
reduce the probability of death for premature babies.

We compare our analysis to several ``baseline'' methods commonly used to
analyze observational studies that are not designed to allow for
unmeasured \mbox{confounders} or nonignorable missingness. The first method we
consider is an unadjusted analysis using the observed rates of neonatal
death in high-level NICUs and low-level NICUs to estimate $\EE(Y\mid
D=1)-\EE(Y\mid D=0)$. The estimate is 0.01 with a $95\%$ confidence
interval [0.009, 0.011], which shows that high-level NICU is associated
with a higher probability of death. The second method we consider is a
logistic regression model of neonatal death indicator $Y$ on treatment~$D$
as well as the measured confounders to get an estimate $\frac{1}{\NN}
\sum_{i=1}^{\NN} [\hat{\EE}(Y\mid D=1, \mathbf{X})-\hat{\EE}(Y\mid D=0,
\mathbf{X})]$ to adjust for covariates. We use mice under the MAR
assumption to impute the missing values in the data. This adjusted
estimate is 0.000, with a $95\%$ confidence interval [$-$0.001, 0.001],
which provides no evidence of an association between level of NICU and
chance of neonatal death. The third method we consider is
subclassification on the propensity score following \citet{RosRub84}. As suggested in \citet{RosRub84}, we divided babies
into five subclasses based on the propensity score, and obtained the
average treatment effect by weighting each subpopulation's average
treatment effect by the proportion of each subclass. This adjusted
analysis shows that the high-level NICU increases the probability of
neonatal death by 0.002, with a 95\% confidence interval [0.001,
0.003]. The conclusions of all the three baseline methods contradicts
with the result of our method, which found evidence that delivery at a
high-level NICU increases a premature baby's chance of survival. Unlike
the three baseline methods, our method allows for unmeasured
confounders and a certain type of nonignorable missingness of covariates.

\section{Sensitivity analysis}\label{sec6}
In this section we will assess the sensitivity of our causal
conclusions to an unmeasured patient risk characteristic relevant to
both the outcome of death and missingness of covariates, for example,
results of tests like fetal heart tracing or doctor's knowledge about
mother's severity of condition. Following the idea of \citet{RosRub}, we assume that there is an unobserved binary covariate $Q$
which represents the risk not explained by compliance status and
gestational age, and that is independent of the observed covariates,
the compliance status and the instrument. We want to know after
accounting for such an unmeasured covariate if there is still evidence
that the high-level NICU reduces the probability of death for babies of
small gestational age.

The adjusted model is as follows:
\[
P(Q=1)=\pi,
\]
the parameter $\pi$ gives the probability that the unobserved binary
risk variable is~1. We assume that the unobserved binary risk variable
$Q$ is independent of IV $Z$, compliance class $U$ and covariates
$\mathbf{X}$, thus, the models (\ref{equ2})--(\ref{equ7}) remain the same in our sensitivity
analysis. The model of outcome controlling also for $Q$ is
\[
P\bigl(Y_i(z)=1\mid U_i=u, \mathbf{X}_i=
\mathbf{x}, Q_i=q\bigr)=\frac
{\exp
{(\beta_{uz}^{T}\mathbf{x}}+\xi_{uz}q)}{1+\exp{(\beta
_{uz}^{T}\mathbf{x}+\xi_{uz}q)}}.
\]
Again, according to Assumption~\ref{ass4}, $\beta_{a0}=\beta_{a1}, \beta
_{n0}=\beta_{n1}, \xi_{a0}=\xi_{a1}$ and $\xi_{n0}=\xi_{n1}$.
$\xi
_{uz}$ gives the log odds ratio for $Y$ in two subpopulations $q=1$ and
$q=0$. Finally, the model for missing indicators of covariate $j$
controlling further for $Q$ is
\begin{eqnarray*}
&& P\bigl(R^{X_{i,j}}_i(z)=1\mid Y_i(z)=y,
U_i=u, \mathbf{X}_{i,1,\ldots,k}=\mathbf{x}_{1,\ldots,k},Q_i=q
\bigr)
\\
&&\qquad =\frac{\exp
{(\theta_{j,u}^T\mathbf{x}_{1,\ldots,k}+\gamma_{j,u}I_{y=1}+\eta
_{j,u}I_{z=1}+\kappa_{j,u}q)}}{1+\exp{(\theta_{j,u}^T\mathbf{x}_{1,\ldots,k}+\gamma_{j,u}I_{y=1}+\eta_{j,u}I_{z=1}+\kappa
_{j,u}q)}},
\end{eqnarray*}
where $ j=k+1,\ldots,m$. Based on Assumption~\ref{ass6}, $\eta_{j,a}=\eta_{j,n}=0$,
$\forall j=k+1,\ldots,m$. $\kappa_{j,u}$ gives the log odds ratio for $R$
in two subpopulations $q=1$ and $q=0$.

For fixed sensitivity parameters $\pi, \xi_{uz}, \kappa_{j,u}$, there
exist unique MLEs of the remaining parameters. Our EM algorithm for the
original model could be easily extended to obtain those estimates. The
average treatment effect for compliers of each covariate level is
estimated by $\mathrm{E}(Y(1)-Y(0)\mid U=c, \mathbf{X}=\mathbf{x}) =\pi\cdot(\frac{1}{1+\exp{(\beta_{c0}^{T}\mathbf{x}+\xi
_{c0})}}-\frac{1}{1+\exp{(\beta_{c1}^{T}\mathbf{x}+\xi
_{c1})}})+(1-\pi)\cdot(\frac{1}{1+\exp{(\beta_{c0}^{T}\mathbf{x})}}-\frac{1}{1+\exp{(\beta_{c1}^{T}\mathbf{x})}})$.

In order to limit the size of the sensitivity analysis, $(\kappa, \xi)$
is assumed in the sensitivity analysis to be the same across all
subclasses defined by IV, compliance class and covariates. And also as
in Table~\ref{tab4}, we estimated CACE for some typical combinations of the
measured confounders under each assignment of $(\pi, \kappa, \xi)$. The
details are presented in Tables~1, 2 and 3 in the supplementary
material [\citet{YanLorSma}], where Table~1 describes the
result for babies of gestational age 24 weeks, Table~2 describes the
result for babies of gestational age 30 weeks and Table~3 describes the
result for babies of gestational age 37 weeks.

%
\begin{table}
\tabcolsep=0pt
\caption{Effects of $Q$ on the CACE for patients with prenatal care
starting at second month of pregnancy, mother's education being high
school and with gestational age being 24 weeks, 30 weeks and 37~weeks,
respectively}\label{tab5}
\begin{tabular*}{\tablewidth}{@{\extracolsep{\fill}}@{}lcc d{2.3} d{2.3} d{2.3}@{}}
\hline
\multirow{2}{44pt}{\textbf{Gestational age}} & \textbf{Effect of $\bolds{Q}$ on $\bolds{Y}$} & \textbf{Effect of $\bolds{Q}$ on $\bolds{R}$} &\multicolumn{3}{c}{$\bolds{P(Q=1)\dvtx \pi}$}\\[-6pt]
& & &\multicolumn{3}{c}{\hrulefill}\\
&&& \multicolumn{1}{c}{\textbf{0.1}} & \multicolumn{1}{c}{\textbf{0.5}} & \multicolumn{1}{c}{\textbf{0.9}}\\
\hline
24& $\exp(\xi)=2$& $\exp(\kappa)=2$& -0.289& -0.283&-0.290\\ 
&& $\exp(\kappa)=\frac{1}{2}$&-0.297 &-0.296 &-0.293\\[4pt]  
& $\exp(\xi)=\frac{1}{2}$& $\exp(\kappa)=2$& -0.293 &-0.296& -0.297\\ 
&& $\exp(\kappa)=\frac{1}{2}$&-0.290& -0.283& -0.289\\[4pt]  
& $\exp(\xi)=3$& $\exp(\kappa)=3$& -0.273 &-0.228 &-0.217\\ 
&& $\exp(\kappa)=\frac{1}{3}$&-0.296 &-0.252 &-0.225\\[4pt]  
& $\exp(\xi)=\frac{1}{3}$& $\exp(\kappa)=3$& -0.298 &-0.329 &-0.379\\ 
&& $\exp(\kappa)=\frac{1}{3}$&-0.289 &-0.300 &-0.352\\ [7pt]
30& $\exp(\xi)=2$& $\exp(\kappa)=2$& -0.032& -0.031& -0.031\\ 
&& $\exp(\kappa)=\frac{1}{2}$& -0.032& -0.033 &-0.032\\[4pt]  
& $\exp(\xi)=\frac{1}{2}$& $\exp(\kappa)=2$& -0.032 &-0.033 &-0.032 \\ 
&& $\exp(\kappa)=\frac{1}{2}$&-0.031& -0.031 &-0.032 \\[4pt]  
& $\exp(\xi)=3$& $\exp(\kappa)=3$& -0.029 &-0.024 &-0.022 \\ 
&& $\exp(\kappa)=\frac{1}{3}$&-0.032& -0.026& -0.022 \\[4pt]  
& $\exp(\xi)=\frac{1}{3}$& $\exp(\kappa)=3$& -0.032& -0.038& -0.046\\ 
&& $\exp(\kappa)=\frac{1}{3}$&-0.032 &-0.035& -0.043 \\ [7pt]
37& $\exp(\xi)=2$& $\exp(\kappa)=2$&0.001& 0.001 & 0.001\\ 
&& $\exp(\kappa)=\frac{1}{2}$&0.001& 0.001 & 0.001\\[4pt]  
& $\exp(\xi)=\frac{1}{2}$& $\exp(\kappa)=2$& 0.001& 0.001& 0.001\\ 
&& $\exp(\kappa)=\frac{1}{2}$&0.001 & 0.001 & 0.001 \\[4pt]  
& $\exp(\xi)=3$& $\exp(\kappa)=3$& 0.001& 0.001 & 0.001 \\ 
&& $\exp(\kappa)=\frac{1}{3}$&0.001 & 0.001& 0.001 \\[4pt]  
& $\exp(\xi)=\frac{1}{3}$& $\exp(\kappa)=3$&0.001& 0.001 & 0.001\\ 
&& $\exp(\kappa)=\frac{1}{3}$&0.001 & 0.001 & 0.001 \\
\hline
\end{tabular*}
\end{table}

Table~\ref{tab5} presents part of the sensitivity analysis results, showing how
the unobserved binary covariate $Q$ affects the CACE for patients with
prenatal care starting at second month of pregnancy, mother's education
being high school and babies' gestational age being 24 weeks, 30 weeks
and 37 weeks, respectively. From Table~\ref{tab5}, we observe that when the odds
ratios are doubled, the estimated CACEs do not change much in each
assignment of sensitivity parameters; and when the odds ratios are
tripled, the estimated CACEs vary more. The same phenomenon could be
observed for other cases in Tables~1--3 in the supplementary material [\citet{YanLorSma}].
It is time consuming to conduct a bootstrap for each combination of
sensitivity parameters to obtain the 95\% confidence interval for each
scenario, however, due to the fact that we are using the same data set
in outcome analysis in Section~\ref{sec5} and also in our sensitivity analysis,
it is reasonable to assume that the width of the confidence intervals
would be similar to the ones shown in Table~\ref{tab4} for each scenario.
Specifically, if the point estimate and the confidence interval for a
parameter in Table~\ref{tab4} is $a$ and [$b$, $c$], respectively, and the point
estimate for a corresponding parameter in the sensitivity analysis
tables (Tables~1, 2 and 3 in the supplementary material) is $d$, then we
estimate the confidence interval for the parameter in the sensitivity
analysis to be $[d-(a-b), d+(c-a)]$. For example, in the first case in
Table~\ref{tab5}, where the gestational age is 24, precare is 2 and mother's
education level is high school, if 10\% of patients' unobserved risk
covariate is 1, and the unobserved covariate doubles both odds ratios
for $Y$ and missingness indicators $R$, the estimated CACE is $-$0.289, with
approximate 95\% confidence interval [$-$0.422, $-$0.130]. We checked the
95\% confidence intervals constructed as above for each case listed in
Tables~1--3 in the supplementary material and find that no confidence
intervals cover 0 for cases shown in Tables~\ref{tab1} (gestational age being
24) and~\ref{tab2} (gestational age being 30), and all confidence intervals
cover 0 for cases in Table~\ref{tab3} (gestational age being 37). Consequently,
the unobserved covariate $Q$ would have to more than triple the odds in
both the outcome and missing indicator models, before altering the
conclusion obtained in Section~\ref{sec5} that high-level NICUs reduce the
probability of death in babies of small gestational age. To provide
some idea about how large an effect an unobserved covariate would have
to be to change our conclusions, we compare the effect to that of the
observed covariate gestational age, which is a strong predictor for
death and risk of complications. According to the fit of our model (see
Table~\ref{tab3}), if gestational age is changed by 2 weeks, then the odds
ratios for the outcome death would be altered by a factor of 2.2 and
the odds ratios for the response would be altered by a factor of 1.6.
Thus, based on our sensitivity analysis results, an unobserved
covariate with the same effect as changing gestational age by 2 weeks
would not change our conclusion that high-level NICUs reduce the
probability of death in babies of small gestational age. We conclude
that even if some confounders, for instance, results of tests like
fetal heart tracing and doctor's knowledge about mother's severity of
condition, are unmeasured and affect both the outcome and missingness
of covariates, they would not change our conclusions unless they had
very large effects.

\section{Summary}\label{sec7}

We proposed a series of models to estimate the causal effect of a
treatment using an instrumental variable when the missingness of
covariates may depend on the fully observed outcome, fully observed
covariates, IV as well as the partially observed compliance behavior.
Simulation studies show that under our nonignorable missingness
assumption where the missingness depends on partially observed
compliance class, even if the missing rate of covariate is low ($12\%
$), and the effect of compliance class on the missingness is only
moderate, the commonly used estimation methods, complete-case analysis
and multiple imputation by chained equations assuming MAR could provide
substantially biased estimates; in contrast, our proposed method, which
is designed to deal with nonignorable missingness of covariates,
provides unbiased results.

In this paper we have developed a maximum likelihood method for
instrumental variable estimation with nonignorable missingness of
covariates. Further research could consider a Bayesian version of our
model which would enable carrying our multiple imputation based on our model.

We applied our method to an observational study of neonatal care that
aims to estimate the delivery effect on mortality of premature babies
being delivered in a high-level NICU vs. a low-level NICU. We found
that high-level NICUs substantially reduce the death risk for babies
with small gestational age, which implies that high-level NICUs are
truly providing considerably better care for babies with small
gestational age. Therefore, it is valuable to invest resources to
strengthen the perinatal regionalization system for those babies. For
babies that are almost mature, strengthening the perinatal
regionalization system should probably not be a priority.

The methods we develop in this paper may be useful for many other
observational studies facing unmeasured confounders as well as
nonignorable missing data like ours. One example we described in the
\hyperref[sec1]{Introduction} is comparative effectiveness studies where it is a concern
that the missingness of important lab values might be related with
compliance status. For these settings, our simulation study shows that
it is important and necessary to model the effect of compliance status
on missingness to get valid estimates.

In this study, we focus on cases which contain missing covariates, and
the missingness of covariates is nonignorable. However, in practice,
many studies face the issue of not only missing covariates but also
missing outcomes. In our nonignorable missingness assumption
(Assumption~\ref{ass6}), we allow the missingness of covariates to depend on the
outcome. If there are also missing outcomes, since the covariates are
predictors for the outcome, it is likely that the missingness of the
outcome is related to the values of covariates which are unobserved for
some subjects. If missingness exists in both the covariates and the
outcome, identifiability is a major issue to study since the
missingness of the covariates and outcome may depend on each other.
Additional assumptions beyond what we have considered are needed for
identifiability. Possible assumptions could be developed based on \citet{PenLitRag04} where missingness of outcome is allowed
to depend on compliance and fully observed data, whereas missingness of
covariates is allowed to depend on only the fully observed data but not
compliance status.

\begin{appendix}
\section*{Appendix: E-step estimates}\label{app}

The fomulas to update $\NN$ in the E-step are as follows, $\forall x_2,x_3,x_4,y$:
{\fontsize{10.45}{12}\selectfont{\begin{eqnarray*}
\hspace*{-5pt}&&\NN_{1,1,x_2,x_3,x_4,a,0,y} = \NNN_{x_2,x_3,x_4,1,0,y},
\\[1pt]
\hspace*{-5pt}&&\NN_{1,1,x_2,x_3,x_4,n,1,y} = \NNN_{x_2,x_3,x_4,0,1,y},
\\[1pt]
\hspace*{-5pt}&&\mathrm{E}\NN_{1,1,x_2,x_3,x_4,a,1,y} = \NNN_{x_2,x_3,x_4,1,1,y}\frac
{P_{1,1,x_2,x_3,x_4,a,1,y}}{P_{1,1,x_2,x_3,x_4,a,1,y}+P_{1,1,x_2,x_3,x_4,c,1,y}},
\\[1pt]
\hspace*{-5pt}&&\mathrm{E}\NN_{1,1,x_2,x_3,x_4,c,1,y} = \NNN_{x_2,x_3,x_4,1,1,y}\frac
{P_{1,1,x_2,x_3,x_4,c,1,y}}{P_{1,1,x_2,x_3,x_4,a,1,y}+P_{1,1,x_2,x_3,x_4,c,1,y}},
\\[1pt]
\hspace*{-5pt}&&\mathrm{E}\NN_{1,1,x_2,x_3,x_4,n,0,y} = \NNN_{x_2,x_3,x_4,0,0,y}\frac
{P_{1,1,x_2,x_3,x_4,
n,0,y}}{P_{1,1,x_2,x_3,x_4,n,0,y}+P_{1,1,x_2,x_3,x_4,c,0,y}},
\\[1pt]
\hspace*{-5pt}&&\mathrm{E}\NN_{1,1,x_2,x_3,x_4,c,0,y} = \NNN_{x_2,x_3,x_4,0,0,y}\frac
{P_{1,1,x_2,x_3,x_4,c,0,y}}{P_{1,1,x_2,x_3,x_4,n,0,y}+P_{1,1,x_2,x_3,x_4,c,0,y}},
\\[1pt]
\hspace*{-5pt}&&\mathrm{E}\NN_{0,1,x_2,x_3,x_4,a,0,y} = \NN3_{x_2,x_4,1,0,y}\frac
{P_{0,1,x_2,x_3,x_4,a,0,y}}{\sum_{x_3=1}^{x_3=q_3}P_{0,1,x_2,x_3,x_4,a,0,y}},
\\[1pt]
\hspace*{-5pt}&&\mathrm{E}\NN_{0,1,x_2,x_3,x_4,n,1,y} = \NN3_{x_2,x_4,0,1,y}\frac
{P_{0,1,x_2,x_3,x_4,n,1,y}}{\sum_{x_3=1}^{x_3=q_3}P_{0,1,x_2,x_3,x_4,n,1,y}},
\\[1pt]
\hspace*{-5pt}&&\mathrm{E}\NN_{0,1,x_2,x_3,x_4,a,1,y}
\\[1pt]
\hspace*{-5pt}&&\qquad\hspace*{-2pt}  = \NN3_{x_2,x_4,1,1,y}\frac
{P_{0,1,x_2,x_3,x_4,a,1,y}}{\sum_{x_3=1}^{x_3=q_3}P_{0,1,x_2,x_3,x_4,a,1,y}+\sum_{x_3=1}^{x_3=q_3}P_{0,1,x_2,x_3,x_4,c,1,y}},
\\[1pt]
\hspace*{-5pt}&&\mathrm{E}\NN_{0,1,x_2,x_3,x_4,c,1,y}
\\[1pt]
\hspace*{-5pt}&&\qquad\hspace*{-2pt} = \NN3_{x_2,x_4,1,1,y}\frac
{P_{0,1,x_2,x_3,x_4,c,1,y}}{\sum_{x_3=1}^{x_3=q_3}P_{0,1,x_2,x_3,x_4,a,1,y}+\sum_{x_3=1}^{x_3=q_3}P_{0,1,x_2,x_3,x_4,c,1,y}},
\\[1pt]
\hspace*{-5pt}&&\mathrm{E}\NN_{0,1,x_2,x_3,x_4,n,0,y}
\\[1pt]
\hspace*{-5pt}&&\qquad\hspace*{-2pt}  = \NN3_{x_2,x_4,0,0,y}\frac
{P_{0,1,x_2,x_3,x_4,n,0,y}}{\sum_{x_3=1}^{x_3=q_3}P_{0,1,x_2,x_3,x_4,n,0,y}+\sum_{x_3=1}^{x_3=q_3}P_{0,1,x_2,x_3,x_4,c,0,y}},
\\[1pt]
\hspace*{-5pt}&&\mathrm{E}\NN_{0,1,x_2,x_3,x_4,c,0,y}
\\[1pt]
\hspace*{-5pt}&&\qquad\hspace*{-2pt}  = \NN3_{x_2,x_4,0,0,y}\frac
{P_{0,1,x_2,x_3,x_4,c,0,y}}{\sum_{x_3=1}^{x_3=q_3}P_{0,1,x_2,x_3,x_4,n,0,y}+\sum_{x_3=1}^{x_3=q_3}P_{0,1,x_2,x_3,x_4,c,0,y}},
\\[1pt]
\hspace*{-5pt}&&\mathrm{E}\NN_{1,0,x_2,x_3,x_4,a,0,y} = \NN4_{x_2,x_3,1,0,y}\frac
{P_{1,0,x_2,x_3,x_4,a,0,y}}{\sum_{x_4=1}^{x_4=q_4}P_{1,0,x_2,x_3,x_4,a,0,y}},
\\[1pt]
\hspace*{-5pt}&&\mathrm{E}\NN_{1,0,x_2,x_3,x_4,n,1,y} = \NN4_{x_2,x_3,0,1,y}\frac
{P_{1,0,x_2,x_3,x_4,n,1,y}}{\sum_{x_4=1}^{x_4=q_4}P_{1,0,x_2,x_3,x_4,n,1,y}},
\\[0.5pt]
\hspace*{-5pt}&&\mathrm{E}\NN_{1,0,x_2,x_3,x_4,a,1,y}
\\[0.5pt]
\hspace*{-5pt}&&\qquad\hspace*{-2pt}  = \NN4_{x_2,x_3,1,1,y}\frac
{P_{1,0,x_2,x_3,x_4,a,1,y}}{\sum_{x_4=1}^{x_4=q_4}P_{1,0,x_2,x_3,x_4,a,1,y}+\sum_{x_4=1}^{x_4=q_4}P_{1,0,x_2,x_3,x_4,c,1,y}},
\\[0.5pt]
\hspace*{-5pt}&&\mathrm{E}\NN_{1,0,x_2,x_3,x_4,c,1,y}
\\[0.5pt]
\hspace*{-5pt}&&\qquad\hspace*{-2pt}  = \NN4_{x_2,x_3,1,1,y}\frac
{P_{1,0,x_2,x_3,x_4,c,1,y}}{\sum_{x_4=1}^{x_4=q_4}P_{1,0,x_2,x_3,x_4,a,1,y}+\sum_{x_4=1}^{x_4=q_4}P_{1,0,x_2,x_3,x_4,c,1,y}},
\\[0.5pt]
\hspace*{-5pt}&&\mathrm{E}\NN_{1,0,x_2,x_3,x_4,n,0,y}
\\[0.5pt]
\hspace*{-5pt}&&\qquad\hspace*{-2pt}  = \NN4_{x_2,x_3,0,0,y}\frac
{P_{1,0,x_2,x_3,x_4,n,0,y}}{\sum_{x_4=1}^{x_4=q_4}P_{1,0,x_2,x_3,x_4,n,0,y}+\sum_{x_4=1}^{x_4=q_4}P_{1,0,x_2,x_3,x_4,c,0,y}},
\\[0.5pt]
\hspace*{-5pt}&&\mathrm{E}\NN_{1,0,x_2,x_3,x_4,c,0,y}
\\[0.5pt]
\hspace*{-5pt}&&\qquad\hspace*{-2pt}  = \NN4_{x_2,x_3,0,0,y}\frac
{P_{1,0,x_2,x_3,x_4,c,0,y}}{\sum_{x_4=1}^{x_4=q_4}P_{1,0,x_2,x_3,x_4,n,0,y}+\sum_{x_4=1}^{x_4=q_4}P_{1,0,x_2,x_3,x_4,c,0,y}},
\\[0.5pt]
\hspace*{-5pt}&&\mathrm{E}\NN_{0,0,x_2,x_3,x_4,a,0,y}
\\[0.5pt]
\hspace*{-5pt}&&\qquad\hspace*{-2pt}  = NB_{x_2,1,0,y}\frac
{P_{0,0,x_2,x_3,x_4,a,0,y}}{\sum_{x_4=1}^{x_4=q_4}\sum_{x_3=1}^{x_3=q_3}P_{1,0,x_2,x_3,x_4,a,0,y}},
\\[0.5pt]
\hspace*{-5pt}&&\mathrm{E}\NN_{0,0,x_2,x_3,x_4,n,1,y}
\\[0.5pt]
\hspace*{-5pt}&&\qquad\hspace*{-2pt}  = NB_{x_2,1,0,y}\frac
{P_{0,0,x_2,x_3,x_4,n,1,y}}{\sum_{x_4=1}^{x_4=q_4}\sum_{x_3=1}^{x_3=q_3}P_{1,0,x_2,x_3,x_4,n,1,y}},
\\[0.5pt]
\hspace*{-5pt}&&\mathrm{E}\NN_{0,0,x_2,x_3,x_4,a,1,y}
\\[0.5pt]
\hspace*{-5pt}&&\qquad\hspace*{-2pt}  = NB_{x_2,1,1,y}\frac
{P_{0,0,x_2,x_3,x_4,a,1,y}}{\sum_{x_4=1}^{x_4=q_4}\sum_{x_3=1}^{x_3=q_3}P_{0,0,x_2,x_3,x_4,a,1,y}+\sum_{x_4=1}^{x_4=q_4}\sum_{x_3=1}^{x_3=q_3}P_{0,0,x_2,x_3,x_4,c,1,y}},
\\[0.5pt]
\hspace*{-5pt}&&\mathrm{E}\NN_{0,0,x_2,x_3,x_4,c,1,y}
\\[0.5pt]
\hspace*{-5pt}&&\qquad\hspace*{-2pt}  = NB_{x_2,1,1,y}\frac
{P_{0,0,x_2,x_3,x_4,c,1,y}}{\sum_{x_4=1}^{x_4=q_4}\sum_{x_3=1}^{x_3=q_3}P_{0,0,x_2,x_3,x_4,a,1,y}+\sum_{x_4=1}^{x_4=q_4}\sum_{x_3=1}^{x_3=q_3}P_{0,0,x_2,x_3,x_4,c,1,y}},
\\[0.5pt]
\hspace*{-5pt}&&\mathrm{E}\NN_{0,0,x_2,x_3,x_4,n,0,y}
\\[0.5pt]
\hspace*{-5pt}&&\qquad\hspace*{-2pt}  = NB_{x_2,1,1,y}\frac
{P_{0,0,x_2,x_3,x_4,n,0,y}}{\sum_{x_4=1}^{x_4=q_4}\sum_{x_3=1}^{x_3=q_3}P_{0,0,x_2,x_3,x_4,n,0,y}+\sum_{x_4=1}^{x_4=q_4}\sum_{x_3=1}^{x_3=q_3}P_{0,0,x_2,x_3,x_4,c,0,y}},
\\[0.5pt]
\hspace*{-5pt}&&\mathrm{E}\NN_{0,0,x_2,x_3,x_4,n,0,y}
\\[0.5pt]
\hspace*{-5pt}&&\qquad\hspace*{-2pt}  = NB_{x_2,1,1,y}\frac
{P_{0,0,x_2,x_3,x_4,c,0,y}}{\sum_{x_4=1}^{x_4=q_4}\sum_{x_3=1}^{x_3=q_3}P_{0,0,x_2,x_3,x_4,n,0,y}+\sum_{x_4=1}^{x_4=q_4}\sum_{x_3=1}^{x_3=q_3}P_{0,0,x_2,x_3,x_4,c,0,y}}.
\end{eqnarray*}}}%
\end{appendix}

\section*{Acknowledgment}
We thank Roland Ramsahai for helpful discussion.


\begin{supplement}
\stitle{Supplement to ``Estimation of causal effects using instrumental variables with
nonignorable missing covariates: Application to effect of type of
delivery NICU on~premature infants''}
\slink[doi]{10.1214/13-AOAS699SUPP} 
\sdatatype{.zip}
\sfilename{aoas699\_supp.zip}
\sdescription{We include in the supplementary document the $R$ code for
the algorithm to analyze our data, discussion on identifiability in the
simplest setup where there is only one covariate which is binary under
both our nonignorable missingness assumption and an alternative
nonignorable missingness assumption, and detailed results of our
sensitivity analysis.}
\end{supplement}


\printaddresses


\begin{thebibliography}{40}

\bibitem[\protect\citeauthoryear{Angrist, Imbens and Rubin}{1996}]{AngImbRub96}
\begin{barticle}[auto:STB|2014/01/06|10:16:28]
\bauthor{\bsnm{Angrist},~\bfnm{J.~D.}\binits{J.~D.}},
\bauthor{\bsnm{Imbens},~\bfnm{G.~W.}\binits{G.~W.}} \AND
\bauthor{\bsnm{Rubin},~\bfnm{D.~B.}\binits{D.~B.}}
(\byear{1996}).
\btitle{Identification of causal effects using instrumental variables}.
\bjournal{J. Amer. Statist. Assoc.}
\bvolume{91}
\bpages{444--455}.
\end{barticle}
\bptok{imsref}%
\endbibitem

\bibitem[\protect\citeauthoryear{Angrist and Krueger}{1991}]{AngKru91}
\begin{barticle}[auto:STB|2014/01/06|10:16:28]
\bauthor{\bsnm{Angrist},~\bfnm{J.~D.}\binits{J.~D.}} \AND
\bauthor{\bsnm{Krueger},~\bfnm{A.~B.}\binits{A.~B.}}
(\byear{1991}).
\btitle{Does compulsory school attendance affect schooling and earnings?}
\bjournal{Quarterly Journal of Economics}
\bvolume{106}
\bpages{979--1014}.
\end{barticle}
\bptok{imsref}%
\endbibitem

\bibitem[\protect\citeauthoryear{Baiocchi et~al.}{2010}]{Baietal10}
\begin{barticle}[mr]
\bauthor{\bsnm{Baiocchi},~\bfnm{Mike}\binits{M.}},
\bauthor{\bsnm{Small},~\bfnm{Dylan~S.}\binits{D.~S.}},
\bauthor{\bsnm{Lorch},~\bfnm{Scott}\binits{S.}} \AND
\bauthor{\bsnm{Rosenbaum},~\bfnm{Paul~R.}\binits{P.~R.}}
(\byear{2010}).
\btitle{Building a stronger instrument in an observational study of perinatal care for premature infants}.
\bjournal{J. Amer. Statist. Assoc.}
\bvolume{105}
\bpages{1285--1296}.
\bid{doi={10.1198/jasa.2010.ap09490}, issn={0162-1459}, mr={2796550}}
\end{barticle}
\bptok{imsref}%
\endbibitem

\bibitem[\protect\citeauthoryear{Boyle et~al.}{1983}]{Boyetal83}
\begin{barticle}[auto:STB|2014/01/06|10:16:28]
\bauthor{\bsnm{Boyle},~\bfnm{M.~H.}\binits{M.~H.}},
\bauthor{\bsnm{Torrance},~\bfnm{G.~W.}\binits{G.~W.}},
\bauthor{\bsnm{Sinclair},~\bfnm{J.~C.}\binits{J.~C.}} \AND
\bauthor{\bsnm{Horwood},~\bfnm{S.~P.}\binits{S.~P.}}
(\byear{1983}).
\btitle{Economic evaluation of neonatal intensive care of very-low-birth-weight infants}.
\bjournal{N.~Engl. J. Med.}
\bvolume{308}
\bpages{1330--13337}.
\end{barticle}
\bptok{imsref}%
\endbibitem

\bibitem[\protect\citeauthoryear{Brookhart and Schneeweiss}{2007}]{BroSch07}
\begin{barticle}[mr]
\bauthor{\bsnm{Brookhart},~\bfnm{M.~Alan}\binits{M.~A.}} \AND
\bauthor{\bsnm{Schneeweiss},~\bfnm{Sebastian}\binits{S.}}
(\byear{2007}).
\btitle{Preference-based instrumental variable methods for the estimation of treatment effects: Assessing validity and interpreting results}.
\bjournal{Int. J. Biostat.}
\bvolume{3}
\bpages{Art. 14, 25}.
\bid{doi={10.2202/1557-4679.1072}, issn={1557-4679}, mr={2383610}}
\end{barticle}
\bptok{imsref}%
\endbibitem

\bibitem[\protect\citeauthoryear{Brookhart et~al.}{2006}]{Broetal06}
\begin{barticle}[auto:STB|2014/01/06|10:16:28]
\bauthor{\bsnm{Brookhart},~\bfnm{M.~A.}\binits{M.~A.}},
\bauthor{\bsnm{Wang},~\bfnm{P.~S.}\binits{P.~S.}},
\bauthor{\bsnm{Solomon},~\bfnm{D.~H.}\binits{D.~H.}} \AND
\bauthor{\bsnm{Schneeweiss},~\bfnm{S.}\binits{S.}}
(\byear{2006}).
\btitle{Evaluating short-term drug effects using a physician-specific prescribing preference as an instrumental variable}.
\bjournal{Epidemiology}
\bvolume{17}
\bpages{268--275}.
\end{barticle}
\bptok{imsref}%
\endbibitem

\bibitem[\protect\citeauthoryear{Chen, Geng and Zhou}{2009}]{CheGenZho09}
\begin{barticle}[mr]
\bauthor{\bsnm{Chen},~\bfnm{Hua}\binits{H.}},
\bauthor{\bsnm{Geng},~\bfnm{Zhi}\binits{Z.}} \AND
\bauthor{\bsnm{Zhou},~\bfnm{Xiao-Hua}\binits{X.-H.}}
(\byear{2009}).
\btitle{Identifiability and estimation of causal effects in randomized trials with noncompliance and completely nonignorable missing data}.
\bjournal{Biometrics}
\bvolume{65}
\bpages{675--682}.
\bid{doi={10.1111/j.1541-0420.2008.01120.x}, issn={0006-341X}, mr={2649840}}
\end{barticle}
\bptok{imsref}%
\endbibitem

\bibitem[\protect\citeauthoryear{Chung et~al.}{2010}]{Chuetal10}
\begin{barticle}[pbm]
\bauthor{\bsnm{Chung},~\bfnm{Judith~H.}\binits{J.~H.}},
\bauthor{\bsnm{Phibbs},~\bfnm{Ciaran~S.}\binits{C.~S.}},
\bauthor{\bsnm{Boscardin},~\bfnm{W.~John}\binits{W.~J.}},
\bauthor{\bsnm{Kominski},~\bfnm{Gerald~F.}\binits{G.~F.}},
\bauthor{\bsnm{Ortega},~\bfnm{Alexander~N.}\binits{A.~N.}} \AND
\bauthor{\bsnm{Needleman},~\bfnm{Jack}\binits{J.}}
(\byear{2010}).
\btitle{The effect of neonatal intensive care level and hospital volume on mortality of very low birth weight infants}.
\bjournal{Med. Care}
\bvolume{48}
\bpages{635--644}.
\bid{doi={10.1097/MLR.0b013e3181dbe887}, issn={1537-1948}, pmid={20548252}}
\end{barticle}
\bptok{imsref}%
\endbibitem


\bibitem[\protect\citeauthoryear{Doyle et~al.}{2004}]{Doy}
\begin{barticle}[auto:STB|2014/01/06|10:16:28]
\bauthor{\bsnm{Doyle},~\bfnm{L.~W.}\binits{L.~W.}}
\AND
\bauthor{Victorian Infant Collaborative Study Group}
(\byear{2004}).
\btitle{Evaluation of neonatal intensive care for extremely low birth weight infants in Victoria over two decades: II. Efficiency}.
\bjournal{Pediatrics}
\bvolume{113}
\bpages{510--514}.
\end{barticle}
\bptok{imsref}%
\endbibitem

\bibitem[\protect\citeauthoryear{Frangakis and Rubin}{1999}]{FraRub99}
\begin{barticle}[mr]
\bauthor{\bsnm{Frangakis},~\bfnm{Constantine~E.}\binits{C.~E.}} \AND
\bauthor{\bsnm{Rubin},~\bfnm{Donald~B.}\binits{D.~B.}}
(\byear{1999}).
\btitle{Addressing complications of intention-to-treat analysis in the combined presence of all-or-none treatment-noncompliance and subsequent missing outcomes}.
\bjournal{Biometrika}
\bvolume{86}
\bpages{365--379}.
\bid{doi={10.1093/biomet/86.2.365}, issn={0006-3444}, mr={1705410}}
\end{barticle}
\bptok{imsref}%
\endbibitem

\bibitem[\protect\citeauthoryear{Guo et al.}{2014}]{GuoChe}
\begin{bmisc}[auto:STB|2014/01/06|10:16:28]
\bauthor{\bsnm{Guo},~\bfnm{Z.}\binits{Z.}},
\bauthor{\bsnm{Cheng},~\bfnm{J.}\binits{J.}},
\bauthor{\bsnm{Lorch},~\bfnm{S.~A.}\binits{S.~A.}} \AND
\bauthor{\bsnm{Small},~\bfnm{D. S.}\binits{D. S.}}
(\byear{2014}).
\bhowpublished{Using an instrumental variable to test for unmeasured confounding. Preprint.}
\end{bmisc}
\bptok{imsref}%
\endbibitem


\bibitem[\protect\citeauthoryear{Howell et~al.}{2002}]{Howetal02}
\begin{barticle}[auto:STB|2014/01/06|10:16:28]
\bauthor{\bsnm{Howell},~\bfnm{E.~M.}\binits{E.~M.}},
\bauthor{\bsnm{Richardson},~\bfnm{D.}\binits{D.}},
\bauthor{\bsnm{Ginsburg},~\bfnm{P.}\binits{P.}} \AND
\bauthor{\bsnm{Foot},~\bfnm{B.}\binits{B.}}
(\byear{2002}).
\btitle{Deregionalization of neonatal intensive care in urban areas}.
\bjournal{Am. J. Publ. Health}
\bvolume{92}
\bpages{119--124}.
\end{barticle}
\bptok{imsref}%
\endbibitem

\bibitem[\protect\citeauthoryear{Imai and Ratkovic}{2013}]{ImaRat13}
\begin{barticle}[mr]
\bauthor{\bsnm{Imai},~\bfnm{Kosuke}\binits{K.}} \AND
\bauthor{\bsnm{Ratkovic},~\bfnm{Marc}\binits{M.}}
(\byear{2013}).
\btitle{Estimating treatment effect heterogeneity in randomized program evaluation}.
\bjournal{Ann. Appl. Stat.}
\bvolume{7}
\bpages{443--470}.
\bid{doi={10.1214/12-AOAS593}, issn={1932-6157}, mr={3086426}}
\bptnote{check year}%
\end{barticle}\vadjust{\goodbreak}
\bptok{imsref}%
\endbibitem

\bibitem[\protect\citeauthoryear{Korn and Baumrind}{1998}]{KorBau98}
\begin{barticle}[mr]
\bauthor{\bsnm{Korn},~\bfnm{Edward~L.}\binits{E.~L.}} \AND
\bauthor{\bsnm{Baumrind},~\bfnm{Sheldon}\binits{S.}}
(\byear{1998}).
\btitle{Clinician preferences and the estimation of causal treatment differences}.
\bjournal{Statist. Sci.}
\bvolume{13}
\bpages{209--235}.
\bid{doi={10.1214/ss/1028905885}, issn={0883-4237}, mr={1665709}}
\bptnote{check related}%
\end{barticle}
\bptok{imsref}%
\endbibitem

\bibitem[\protect\citeauthoryear{Lasswell et~al.}{2010}]{Lasetal10}
\begin{barticle}[auto:STB|2014/01/06|10:16:28]
\bauthor{\bsnm{Lasswell},~\bfnm{S.~M.}\binits{S.~M.}},
\bauthor{\bsnm{Barfield},~\bfnm{W.~D.}\binits{W.~D.}},
\bauthor{\bsnm{Rochat},~\bfnm{R.~W.}\binits{R.~W.}} \AND
\bauthor{\bsnm{Blackmon},~\bfnm{L.}\binits{L.}}
(\byear{2010}).
\btitle{Perinatal regionalization for very low-birth-weight and very preterm infants: A meta-analysis}.
\bjournal{J. Am. Med. Assoc.}
\bvolume{304}
\bpages{992--1000}.
\end{barticle}
\bptok{imsref}%
\endbibitem

\bibitem[\protect\citeauthoryear{Levy, O'Malley and Normand}{2004}]{LevOMaNor04}
\begin{barticle}[auto:STB|2014/01/06|10:16:28]
\bauthor{\bsnm{Levy},~\bfnm{D.~E.}\binits{D.~E.}},
\bauthor{\bsnm{O'Malley},~\bfnm{A.~J.}\binits{A.~J.}} \AND
\bauthor{\bsnm{Normand},~\bfnm{S.~T.}\binits{S.~T.}}
(\byear{2004}).
\btitle{Covariate adjustment in clinical trials with nonignorable missing data and noncompliance}.
\bjournal{Stat. Med.}
\bvolume{23}
\bpages{2319--2339}.
\end{barticle}
\bptok{imsref}%
\endbibitem

\bibitem[\protect\citeauthoryear{Little and Rubin}{2002}]{LitRub02}
\begin{bbook}[mr]
\bauthor{\bsnm{Little},~\bfnm{Roderick~J.~A.}\binits{R.~J.~A.}} \AND
\bauthor{\bsnm{Rubin},~\bfnm{Donald~B.}\binits{D.~B.}}
(\byear{2002}).
\btitle{Statistical Analysis with Missing Data},
\bedition{2nd} ed.
\bpublisher{Wiley},
\blocation{Hoboken, NJ}.
\bid{mr={1925014}}
\end{bbook}
\bptok{imsref}%
\endbibitem

\bibitem[\protect\citeauthoryear{Lorch et~al.}{2012}]{LorBai}
\begin{barticle}[auto:STB|2014/01/06|10:16:28]
\bauthor{\bsnm{Lorch},~\bfnm{S.~A.}\binits{S.~A.}},
\bauthor{\bsnm{Baiocchi},~\bfnm{M.}\binits{M.}},
\bauthor{\bsnm{Ahlberg},~\bfnm{C.~E.}\binits{C.~E.}} \AND
\bauthor{\bsnm{Small},~\bfnm{D.~S.}\binits{D.~S.}}
(\byear{2012}).
\btitle{The differential impact of delivery NICU on the outcomes of premature infance}.
\bjournal{Pediatrics}
\bvolume{130}
\bpages{1--9}.
\end{barticle}
\bptok{imsref}%
\endbibitem

\bibitem[\protect\citeauthoryear{Mealli et~al.}{2004}]{Meaetal04}
\begin{barticle}[auto:STB|2014/01/06|10:16:28]
\bauthor{\bsnm{Mealli},~\bfnm{F.}\binits{F.}},
\bauthor{\bsnm{Imbens},~\bfnm{G.}\binits{G.}},
\bauthor{\bsnm{Ferro},~\bfnm{S.}\binits{S.}} \AND
\bauthor{\bsnm{Biggeri},~\bfnm{A.}\binits{A.}}
(\byear{2004}).
\btitle{Analyzing a randomized trial on breast self examination with noncompliance and missing outcomes}.
\bjournal{Biostatistics}
\bvolume{5}
\bpages{207--222}.
\end{barticle}
\bptok{imsref}%
\endbibitem

\bibitem[\protect\citeauthoryear{Olkin and Tate}{1961}]{OlkTat61}
\begin{barticle}[mr]
\bauthor{\bsnm{Olkin},~\bfnm{I.}\binits{I.}} \AND
\bauthor{\bsnm{Tate},~\bfnm{R.~F.}\binits{R.~F.}}
(\byear{1961}).
\btitle{Multivariate correlation models with mixed discrete and continuous variables}.
\bjournal{Ann. Inst. Statist. Math.}
\bvolume{32}
\bpages{448--465}.
\bid{issn={0003-4851}, mr={0152062}}
\bptnote{check related}%
\end{barticle}
\bptok{imsref}%
\endbibitem

\bibitem[\protect\citeauthoryear{Peng, Little and Raghunathan}{2004}]{PenLitRag04}
\begin{barticle}[mr]
\bauthor{\bsnm{Peng},~\bfnm{Yahong}\binits{Y.}},
\bauthor{\bsnm{Little},~\bfnm{Roderick~J.~A.}\binits{R.~J.~A.}} \AND
\bauthor{\bsnm{Raghunathan},~\bfnm{Trivellore~E.}\binits{T.~E.}}
(\byear{2004}).
\btitle{An extended general location model for causal inferences from data subject to noncompliance and missing values}.
\bjournal{Biometrics}
\bvolume{60}
\bpages{598--607}.
\bid{doi={10.1111/j.0006-341X.2004.00208.x}, issn={0006-341X}, mr={2089434}}
\end{barticle}
\bptok{imsref}%
\endbibitem

\bibitem[\protect\citeauthoryear{Phibbs et~al.}{1993}]{PhiMarLuf93}
\begin{barticle}[auto:STB|2014/01/06|10:16:28]
\bauthor{\bsnm{Phibbs},~\bfnm{C.~S.}\binits{C.~S.}},
\bauthor{\bsnm{Mark},~\bfnm{D.~H.}\binits{D.~H.}},
\bauthor{\bsnm{Luft},~\bfnm{H.~S.}\binits{H.~S.}} \betal{et~al.}
(\byear{1993}).
\btitle{Choice of hospital for delivery: A comparison of high-risk and low-risk women}.
\bjournal{Health Serv. Res.}
\bvolume{28}
\bpages{201--222}.
\end{barticle}
\bptok{imsref}%
\endbibitem

\bibitem[\protect\citeauthoryear{Phibbs et~al.}{2007}]{Phietal07}
\begin{barticle}[auto:STB|2014/01/06|10:16:28]
\bauthor{\bsnm{Phibbs},~\bfnm{C.~S.}\binits{C.~S.}},
\bauthor{\bsnm{Baker},~\bfnm{L.~C.}\binits{L.~C.}},
\bauthor{\bsnm{Caughey},~\bfnm{A.~B.}\binits{A.~B.}},
\bauthor{\bsnm{Danielsen},~\bfnm{B.}\binits{B.}},
\bauthor{\bsnm{Schmitt},~\bfnm{S.~K.}\binits{S.~K.}} \AND
\bauthor{\bsnm{Phibbs},~\bfnm{R.~H.}\binits{R.~H.}}
(\byear{2007}).
\btitle{Level and volume of neonatal intensive care and mortality in very-low-birth-weight infants}.
\bjournal{N. Engl. J. Med.}
\bvolume{356}
\bpages{2165--2175}.
\end{barticle}
\bptok{imsref}%
\endbibitem

\bibitem[\protect\citeauthoryear{Profit et~al.}{2010}]{Proetal}
\begin{barticle}[auto:STB|2014/01/06|10:16:28]
\bauthor{\bsnm{Profit},~\bfnm{J.}\binits{J.}},
\bauthor{\bsnm{Lee},~\bfnm{D.}\binits{D.}},
\bauthor{\bsnm{Zupancic},~\bfnm{J.~A.}\binits{J.~A.}},
\bauthor{\bsnm{Papile},~\bfnm{L.}\binits{L.}},
\bauthor{\bsnm{Gutierrez},~\bfnm{C.}\binits{C.}},
\bauthor{\bsnm{Goldie},~\bfnm{S.~J.}\binits{S.~J.}},
\bauthor{\bsnm{Gonzalez-Pier},~\bfnm{E.}\binits{E.}} \AND
\bauthor{\bsnm{Salomon},~\bfnm{J.~A.}\binits{J.~A.}}
(\byear{2010}).
\btitle{Clinical benefits, costs, and cost-effectiveness of neonatal intensive care in Mexico}.
\bjournal{PLoS Medicine}
\bvolume{7}
\bpages{1--10}.
\end{barticle}
\bptok{imsref}%
\endbibitem

\bibitem[\protect\citeauthoryear{Richardson et~al.}{1995}]{RicReeCut95}
\begin{barticle}[auto:STB|2014/01/06|10:16:28]
\bauthor{\bsnm{Richardson},~\bfnm{D.~K.}\binits{D.~K.}},
\bauthor{\bsnm{Reed},~\bfnm{K.}\binits{K.}},
\bauthor{\bsnm{Cutler},~\bfnm{J.~C.}\binits{J.~C.}} \betal{et~al.}
(\byear{1995}).
\btitle{Perinatal regionalization vs hospital competition: The Hartford example}.
\bjournal{Pediatrics}
\bvolume{96}
\bpages{417--423}.
\end{barticle}
\bptok{imsref}%
\endbibitem

\bibitem[\protect\citeauthoryear{Rogowski et~al.}{2004}]{Rogetal04}
\begin{barticle}[auto:STB|2014/01/06|10:16:28]
\bauthor{\bsnm{Rogowski},~\bfnm{J.~A.}\binits{J.~A.}},
\bauthor{\bsnm{Horbar},~\bfnm{J.~D.}\binits{J.~D.}},
\bauthor{\bsnm{Staiger},~\bfnm{D.~O.}\binits{D.~O.}},
\bauthor{\bsnm{Kenny},~\bfnm{M.}\binits{M.}},
\bauthor{\bsnm{Carpenter},~\bfnm{J.}\binits{J.}} \AND
\bauthor{\bsnm{Geppert},~\bfnm{J.}\binits{J.}}
(\byear{2004}).
\btitle{Indirect vs direct hospital quality indicators for very low-birth-weight infants}.
\bjournal{J. Am. Med. Assoc.}
\bvolume{291}
\bpages{202--209}.
\end{barticle}
\bptok{imsref}%
\endbibitem

\bibitem[\protect\citeauthoryear{Rosenbaum and Rubin}{1983}]{RosRub}
\begin{barticle}[auto:STB|2014/01/06|10:16:28]
\bauthor{\bsnm{Rosenbaum},~\bfnm{P.~R.}\binits{P.~R.}} \AND
\bauthor{\bsnm{Rubin},~\bfnm{D.~B.}\binits{D.~B.}}
(\byear{1983}).
\btitle{Assessing sensitivity to an unobserved binary covariate in an observational study with binary outcome}.
\bjournal{J. R. Stat. Soc. Ser. B Stat. Methodol.}
\bvolume{45}
\bpages{212--218}.
\end{barticle}
\bptok{imsref}%
\endbibitem

\bibitem[\protect\citeauthoryear{Rosenbaum and Rubin}{1984}]{RosRub84}
\begin{barticle}[auto:STB|2014/01/06|10:16:28]
\bauthor{\bsnm{Rosenbaum},~\bfnm{P.~R.}\binits{P.~R.}} \AND
\bauthor{\bsnm{Rubin},~\bfnm{D.~B.}\binits{D.~B.}}
(\byear{1984}).
\btitle{Reducing bias in observational studies using subclassification on the propensity score}.
\bjournal{J. Amer. Statist. Assoc.}
\bvolume{79}
\bpages{516--524}.
\end{barticle}
\bptok{imsref}%
\endbibitem

\bibitem[\protect\citeauthoryear{Roy and Hennessy}{2011}]{RoyHen11}
\begin{barticle}[auto:STB|2014/01/06|10:16:28]
\bauthor{\bsnm{Roy},~\bfnm{J.}\binits{J.}} \AND
\bauthor{\bsnm{Hennessy},~\bfnm{S.}\binits{S.}}
(\byear{2011}).
\btitle{Bayesian hierarchical pattern mixture models for comparative effectiveness of drugs and drug classes using healthcare data: A case study involving antihypertensive medications}.
\bjournal{Statistics in Biosciences}
\bvolume{3}
\bpages{79--93}.
\end{barticle}
\bptok{imsref}%
\endbibitem

\bibitem[\protect\citeauthoryear{Schafer}{1997}]{Sch97}
\begin{bbook}[mr]
\bauthor{\bsnm{Schafer},~\bfnm{J.~L.}\binits{J.~L.}}
(\byear{1997}).
\btitle{Analysis of Incomplete Multivariate Data}.
\bpublisher{Chapman \& Hall},
\blocation{London}.
\bid{doi={10.1201/9781439821862}, mr={1692799}}
\end{bbook}
\bptok{imsref}%
\endbibitem

\bibitem[\protect\citeauthoryear{Small and Cheng}{2009}]{SmaChe09}
\begin{barticle}[mr]
\bauthor{\bsnm{Small},~\bfnm{Dylan~S.}\binits{D.~S.}} \AND
\bauthor{\bsnm{Cheng},~\bfnm{Jing}\binits{J.}}
(\byear{2009}).
\btitle{Discussions of
``Identifiability and estimation of causal effects in randomized trials with noncompliance and completely nonignorable missing data.''}
\bjournal{Biometrics}
\bvolume{65}
\bpages{682--686}.
\bid{doi={10.1111/j.1541-0420.2008.01121.x}, issn={0006-341X}, mr={2766612}}
\end{barticle}
\bptok{imsref}%
\endbibitem


\bibitem[\protect\citeauthoryear{Van Buuren and Groothuis-Oudshoorn}{2011}]{VanGro}
\begin{barticle}[auto:STB|2014/01/06|10:16:28]
\bauthor{\bparticle{Van} \bsnm{Buuren},~\bfnm{S.}\binits{S.}} \AND
\bauthor{\bsnm{Groothuis-Oudshoorn},~\bfnm{K.}\binits{K.}}
(\byear{2011}).
\btitle{mice: Multivariate imputation by chained equations in R}.
\bjournal{J. Stat. Softw.}
\bvolume{45}
\bpages{1--67}.
\end{barticle}
\bptok{imsref}%
\endbibitem

\bibitem[\protect\citeauthoryear{Walker}{1996}]{Wal96}
\begin{barticle}[auto:STB|2014/01/06|10:16:28]
\bauthor{\bsnm{Walker},~\bfnm{A.}\binits{A.}}
(\byear{1996}).
\btitle{Confounding by indication}.
\bjournal{Epidemiology}
\bvolume{7}
\bpages{335--336}.
\end{barticle}
\bptok{imsref}%
\endbibitem


\bibitem[\protect\citeauthoryear{Yang, Lorch and Small}{2014}]{YanLorSma}
\begin{bmisc}[auto:STB|2014/01/06|10:16:28]
\bauthor{\bsnm{Yang},~\bfnm{F.}\binits{F.}},
\bauthor{\bsnm{Lorch},~\bfnm{S.~A.}\binits{S.~A.}} \AND
\bauthor{\bsnm{Small},~\bfnm{D.~S.}\binits{D.~S.}}
(\byear{2014}).
\bhowpublished{Supplement to ``Estimation of causal effects using instrumental variables with
nonignorable missing covariates: Application to effect of type of
delivery NICU on~premature infants.''
DOI:\doiurl{10.1214/13-AOAS699SUPP}}.
\bptok{imsref}%
\end{bmisc}
%
\endbibitem



\bibitem[\protect\citeauthoryear{Yeast et~al.}{1998}]{Yeaetal98}
\begin{barticle}[auto:STB|2014/01/06|10:16:28]
\bauthor{\bsnm{Yeast},~\bfnm{J.~D.}\binits{J.~D.}},
\bauthor{\bsnm{Poskin},~\bfnm{M.}\binits{M.}},
\bauthor{\bsnm{Stockbauer},~\bfnm{J.~W.}\binits{J.~W.}} \AND
\bauthor{\bsnm{Shaffer},~\bfnm{S.}\binits{S.}}
(\byear{1998}).
\btitle{Changing patterns in regionalization of perinatal care and the impact on neonatal mortality}.
\bjournal{Am. J. Obstet. Gynecol.}
\bvolume{178}
\bpages{131--135}.
\end{barticle}
\bptok{imsref}%
\endbibitem



\end{thebibliography}
\end{document}